\documentclass[twocolumn,tighten]{aastex62}
\usepackage{amssymb, amsmath,framed}

\newcommand\msun{M$_\odot$}

\newcommand\eg{\citep[eg.][]}

\graphicspath{{./}{figures/}}

\shorttitle{Habitability of Brown Dwarf Exoplanets}
\shortauthors{Lingam et al.}

\begin{document}

\title{Prospects for Life on Temperate Planets Around Brown Dwarfs}

\correspondingauthor{Manasvi Lingam}
\email{manasvi.lingam@cfa.harvard.edu}

\author{Manasvi Lingam}
\affiliation{Institute for Theory and Computation, Harvard University, Cambridge MA 02138, USA}
\affiliation{Department of Aerospace, Physics and Space Sciences, Florida Institute of Technology, Melbourne FL 32901, USA}

\author{Idan Ginsburg}
\affil{Institute for Theory and Computation, Harvard University, Cambridge MA 02138, USA}

\author{Abraham Loeb}
\affiliation{Institute for Theory and Computation, Harvard University, Cambridge MA 02138, USA}

\begin{abstract}
There is growing evidence that brown dwarfs may be comparable to main-sequence stars in terms of their abundance. In this paper, we explore the prospects for the existence of life on Earth-like planets around brown dwarfs. We consider the following factors: (i) the length of time that planets can exist in the temporally shifting habitable zone, (ii) the minimum photon fluxes necessary for oxygenic photosynthesis, and (iii) the lower limits on the fluxes of ultraviolet radiation to drive prebiotic reactions ostensibly necessary for the origin of life. By taking these effects into consideration, we find that it is unlikely for brown dwarfs with masses $\lesssim 30 M_J$ to host habitable planets over geologically significant timescales. We also briefly discuss some of the major biosignatures that might arise on these planets, assess the likelihood of their detection, and highlight some avenues for further study.\\
\end{abstract}

\section{Introduction} \label{sec:intro}

The misnomered ``brown dwarfs'' are substellar objects that straddle the line between hydrogen-burning stars and gas giant planets such as Jupiter. First predicted by \citet{SSK62,Kum62,Kumar63} and \citet{Hayashi63}, it would be more than three decades later before the first brown dwarf was confirmed \citep{Rebolo95}. This brown dwarf, Teide 1, has a spectral classification of M8 \citep{Rebolo14}. Teide 1's designation is based upon the classification system devised by \citet{Kirkpatrick91} which classifies M-dwarfs according to the strength of the TiO and VO bands. 
Thanks to the {\it Two Micron All-Sky Survey} (2MASS),\footnote{\url{https://old.ipac.caltech.edu/2mass/}} the discovery of even cooler brown dwarfs led to the introduction of the ``L" spectral class \citep{Martin97, Kirkpatrick99}. In this classification L-dwarf objects have fewer TiO and VO bands in their spectra, and more metallic hydrides (CrH, FeH) as well as neutral alkali metals. As surveys discovered more brown dwarfs, it became clear that objects such as Gliese 229B \citep{Oppenheimer95} did not belong to either the M or L class \citep{Cushing14}. These cooler objects show methane (CH$_4$) absorption in the near-infrared, and thus the ``T" spectral class was introduced \citep{Kirkpatrick99,Burgasser02,Geballe02,Kirk05}. 

However, different classification schemes led to confusion within the burgeoning brown dwarf community until a unified near-infrared classification scheme for T-dwarfs was adopted \citep{Burgasser06}, whereby the spectral sequence shows increasing H$_2$O and CH$_4$ absorption in later-type T-dwarfs \citep{Lod10}. A few years later, the {\it Wide-field Infrared Survey Exolporer (WISE\footnote{http://wise.ssl.berkeley.edu/})} discovered seven ``ultracool" brown dwarfs, which are believed to be distinct from T-dwarfs and that show ammonia (NH$_3$) features in their spectra; thus the ``Y" spectral class was created \citep{Cushing11,KGC12}. Currently, the coldest known brown dwarfs have an effective temperature of $\sim 250$ K, with one notable example being WISE J085510.83-071442.5 \citep{Luh14}. However, their masses are typically $\lesssim 10 M_J$, where $M_J$ is the mass of Jupiter, and thus their classification is subject to some ambiguity \citep{Cab18}.

Another way to delineate brown dwarfs is via their energy production mechanism. A main-sequence star in hydrostatic equilibrium will fuse hydrogen to helium. On the other hand, a brown dwarf will be below the threshold necessary for core hydrogen burning, yet massive enough to ignite deuterium burning \eg{Burrows93,Burrows01,Spiegel11,Luh12,Auddy16}. This places the mass of a brown dwarf ($M_\mathrm{BD}$) in the range $13 M_J \lesssim M_\mathrm{BD} \lesssim 90 M_J$.\footnote{For the sake of simplicity, we shall truncate our plots in this paper at $70 M_J$ because the smallest ultracool dwarf stars currently detected have masses close to this limit.} However, \citet{Forbes19} demonstrated that it may be possible, in principle, to have brown dwarfs as massive as 0.12\msun\, and yet not reach core hydrogen burning. Clearly, our picture of brown dwarfs is incomplete and will doubtless evolve as we learn more. In this paper, we follow the conventional definition of a brown dwarf as an object that does not reach hydrogen burning but is massive enough for deuterium fusion. 

Recent surveys suggest that there may be as many as $\sim 0.2$-$0.5$ brown dwarfs per star within the Milky Way galaxy \citep{SGC13,Muzic17,Muz19,SDR19} and this ratio might be even higher if the population of cooler Y dwarfs has been insufficiently sampled (see \citealt{KMS19}). Furthermore, the majority of brown dwarfs likely have protoplanetary disks, much the same as Sun-like stars, which can result in the formation of Earth-sized or smaller planets \citep{Apai05,Payne07,Apai13,DNS16}. \citet{Chauvin04} were the first to report a planetary candidate ($\sim 4 M_J$) around a brown dwarf. In the ensuing years, other brown dwarfs with Jupiter-mass planets have been detected \eg{TLM10,Han13,Jung18}. Via microlensing, \citet{Udalski15} seemingly discovered a Venus-mass planet orbiting a brown dwarf, although the data remains open to alternative interpretations that are compatible with its non-existence \citep{HBU16}. 

This naturally raises the question: What are the prospects for habitable planets around brown dwarfs? Despite the importance of this question, comparatively few studies have addressed it. Before proceeding further, it is worth noting that a couple of studies have explored the prospects for life \emph{within} the atmospheres of brown dwarfs, but this topic shall not be considered here \citep{Shap67,Yates17,Lingam19}. Many important aspects of planetary habitability are determined by the properties of the host brown dwarf. A classic example concerns the inner and outer limits of the habitable zone, i.e., the region where the planet can theoretically sustain liquid water on its surface \citep{Dole,Kasting:93,Kopparapu13,Ram18}. While some of the earlier models of the habitable zones around brown dwarfs did not take tidal and geological effects into consideration \citep{Desi99,AS04}, this limitation was addressed by subsequent studies \citep{BRL11,Barnes13,Bol18}. 

The second topic of importance is photosynthesis because it constitutes the primary source of biomass on Earth, and its emergence facilitated the transformation of Earth's geological, chemical and biological landscapes \citep{Knoll15}. The only study that briefly addressed the prospects for photosynthesis on brown dwarfs was by \citet{RaD13}, but the temporal evolution of brown dwarfs was not fully accounted for. Lastly, there is emerging (albeit tentative and disputed) evidence that the origin of life may have necessitated access to sufficient fluxes of UV-C radiation \citep{Suth17,ManLo}. In this scenario, the fluxes incident on temperate planets around brown dwarfs would depend on the properties of the latter. While this issue has been investigated for planets around main-sequence stars \citep{GZZ10,RXT18}, no equivalent studies have been conducted for brown dwarfs. 

Thus, in this paper we examine all of these three topics - the habitable zone, photosynthesis and UV-mediated prebiotic chemistry. The outline of the paper is as follows. We explore some properties of the habitable zones of brown dwarfs in Sec. \ref{SecHZBD} and describe some of the strengths and limitations of our model. In Sec. \ref{SecPHZ}, we explore the constraints on ``Earth-like'' planets around brown dwarfs to sustain oxygenic photosynthesis. We follow this up in Sec. \ref{SecUVAbio} with a study of the equivalent constraints on UV radiation to facilitate abiogenesis on these worlds. Finally, we discuss some of the potential biosignatures and their detectability along with a summary of our findings in Sec. \ref{SecDisc}.

\section{Duration of the habitable zone for brown dwarfs}\label{SecHZBD}
The processes underlying the cooling of brown dwarfs are complex, especially during the early stages of their evolution \citep{Burrows01}. For the purposes of our analysis, we shall employ the formulae presented in \citet{Burrows93}. The effective temperature ($T_\mathrm{BD}$) of the brown dwarf is directly taken from equation (2.58) of \citet{Burrows93}, and has the form
\begin{eqnarray}\label{TeffBD}
 && T_\mathrm{BD} \approx 59\,\mathrm{K}\,\left(\frac{t_\mathrm{BD}}{1\,\mathrm{Gyr}}\right)^{-0.324} \left(\frac{M_\mathrm{BD}}{M_J}\right)^{0.827} \nonumber \\
 && \hspace{0.3in} \times \left(\frac{\kappa_R}{0.01\,\mathrm{cm^2/g}}\right)^{0.088},  
\end{eqnarray}
where $t_\mathrm{BD}$ and $M_\mathrm{BD}$ are the age and mass of the brown dwarf, respectively, while $\kappa_R$ denotes the Rosseland mean opacity of the brown dwarf near its photosphere; recall that $M_J$ is the mass of Jupiter. As the dependence of $T_\mathrm{BD}$ on $\kappa_R$ is weak, we shall neglect the last term on the right-hand-side of the above equation without much loss of generality. Another useful expression in our subsequent analysis is equation (2.37) of \citet{Burrows93} for the radius ($R_\mathrm{BD}$) of the brown dwarf:
\begin{equation}\label{RadBD}
 R_\mathrm{BD} \approx 35.5 R_\oplus\,\left(\frac{M_\mathrm{BD}}{M_J}\right)^{-1/3}.
\end{equation}

In order to assess the length of time that an Earth-like planet spends in the habitable zone of a brown dwarf (denoted by $t_\mathrm{HZ}$), we have to take into account the cooling time and mass of the brown dwarf as well as the planet's orbital radius, $a$. Although we will treat $a$ as being constant over time, it may increase by a factor of $\lesssim 2$ due to outward migration driven by tidal evolution effects \citep{BRL11}. We note that $t_\mathrm{HZ}$ has been investigated by several authors \citep{Desi99,AS04,BRL11,Barnes13}. While our model does not incorporate tidal effects, it has the advantage of being physically transparent and analytically tractable. 

We shall restrict ourselves to planetary systems that satisfy $t_\mathrm{HZ} > 10$ Myr. We impose this lower bound for three reasons. First and foremost, we do so because the cooling ansatz used for brown dwarfs in (\ref{TeffBD}) is not applicable at timescales $\lesssim 10$ Myr \citep{Burrows01}.\footnote{We thank Adam Burrows for clarifying this point.} Second, a timescale of $\sim 1$-$10$ Myr is generally regarded as being necessary for planet formation around brown dwarfs \citep{BRL11,Apai13}. Lastly, from an evolutionary perspective, a timescale of $\sim 10$ Myr is smaller by $\sim 2$ orders of magnitude than the interval required for many of the major evolutionary developments (i.e., critical steps) in Earth's history \citep{Cart08,Knoll15,LL19}, with the caveat that the pace of evolution on other worlds could be substantially faster or slower.

In addition, there is a crucial constraint on $a$ that must be taken into account. At the very minimum, $a$ must not be within the Roche limit, otherwise the planet would be subject to tidal disruption. If the secondary object is fluid-like, the Roche limit ($d_\mathrm{RL}$) for the brown dwarf-planet system is given by \citep{MD99}:
\begin{equation}\label{RocLim}
   d_\mathrm{RL} \approx 2.46 R_\mathrm{BD} \left(\frac{\rho_\mathrm{BD}}{\rho_\mathrm{planet}}\right)^{1/3}, 
\end{equation}
where $\rho_\mathrm{BD}$ and $\rho_\mathrm{planet}$ are the densities of the brown dwarf and the planet, respectively. After further simplification using equation (2.38) of \citet{Burrows93} for $\rho_\mathrm{BD}$ along with the adoption of a mean planetary density approximately equal to the Earth's density of $\rho_\oplus \approx 5.5$ g/cm$^3$, we obtain
\begin{equation}\label{RocFin}
   d_\mathrm{RL} \approx 1.3 \times 10^{-3}\,\mathrm{AU}\, \left(\frac{M_\mathrm{BD}}{M_J}\right)^{1/3}. 
\end{equation}
Alternatively, we can reformulate (\ref{RocLim}) to arrive at
\begin{equation}
    \frac{d_\mathrm{RL}}{R_\mathrm{BD}} \approx 0.86 \left(\frac{M_\mathrm{BD}}{M_J}\right)^{2/3},
\end{equation}
from which it is apparent that the tidal limit is very close to the surface of the brown dwarf. It should therefore be recognized that $d_\mathrm{RL}$ is merely a lower limit on the orbital radius. At small values of $a$, the insolation received can exceed the threshold associated with the inner edge of the habitable zone, as discussed later. Second, at sufficiently close-in distances, the effects of tidal heating become important and may lead to the planet entering a runaway greenhouse state. We do not explicitly tackle this factor in our calculations because the degree of heating depends on a number of parameters such as the eccentricity, inclination and rheology of the planet, as well as the mass and composition of the brown dwarf \citep{Barnes13,Bol18}.

\begin{figure}
\includegraphics[width=7.5cm]{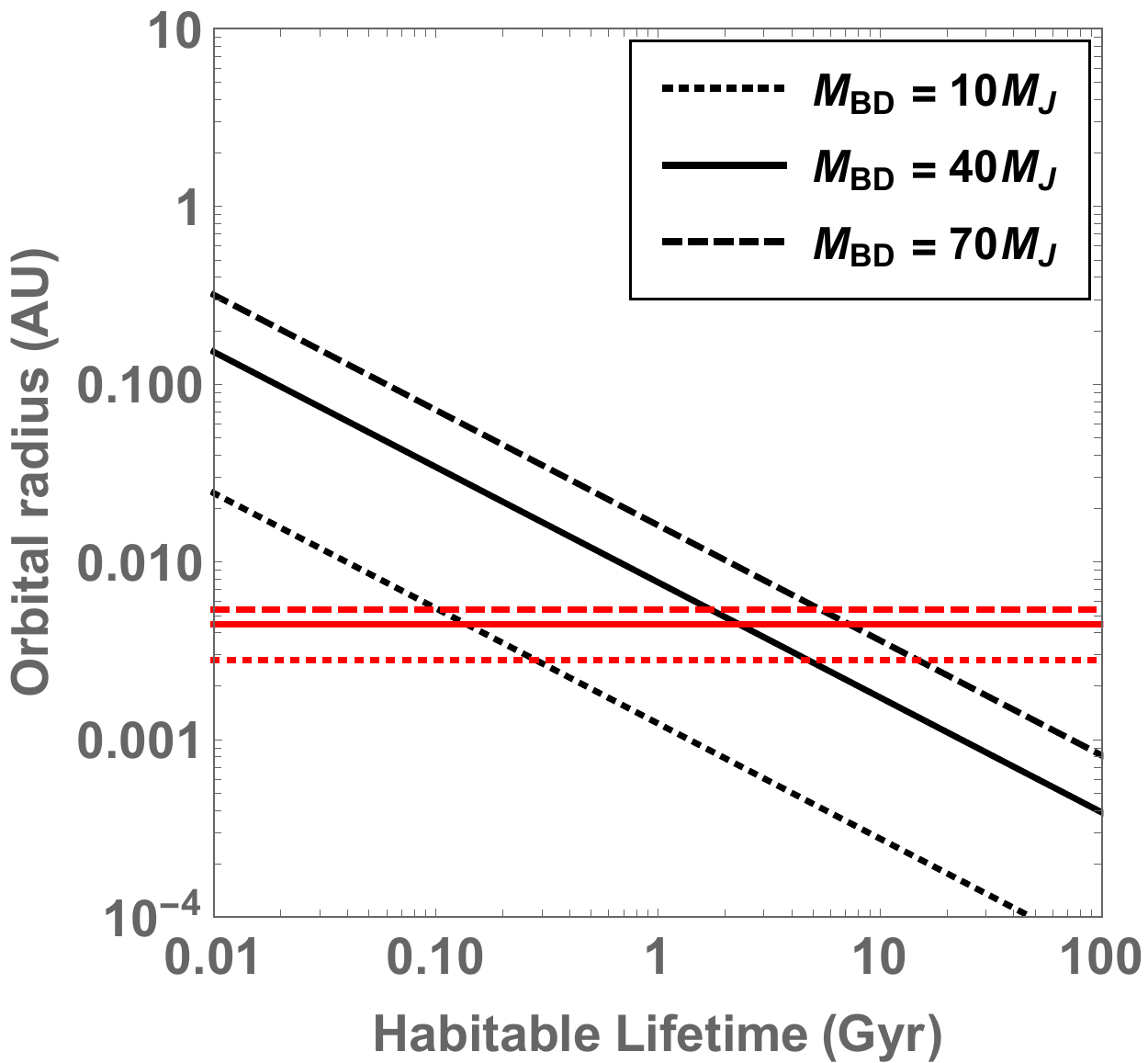} \\
\caption{The orbital radius (in AU) compatible with a given time interval that the planet spends in the habitable zone ($t_\mathrm{HZ}$) is plotted for brown dwarfs of different masses. The dotted, unbroken and dashed black curves correspond to brown dwarf masses of $10 M_J$, $40 M_J$ and $70 M_J$, respectively. The dotted, unbroken and dashed red lines are the Roche limits for $10 M_J$, $40 M_J$ and $70 M_J$ brown dwarfs, respectively.}
\label{FigBDHZ}
\end{figure}

The effective temperature of the planet ($T_P$) can be readily estimated by using
\begin{equation}
   T_P =  T_\mathrm{BD} \sqrt{\frac{R_\mathrm{BD}}{2a}} \left(1 - A_P\right)^{1/4},
\end{equation}
where $A_P$ is the Bond albedo of the planet. Unless the planet has an albedo very close to unity, the last term on the right-hand-side will only affect our results by a factor of $\lesssim 2$. Hence, we shall set $A_P \approx 0.3$ to maintain consistency with the Earth. In actuality, the albedo will depend on a number of factors such as the atmospheric and surface composition of the planet as well as the spectral type of the brown dwarf (or star). In order to determine the length of time that the planet can remain in the habitable zone, it is necessary to choose appropriate upper ($T_\mathrm{max}$) and lower ($T_\mathrm{min}$) bounds on $T_P$. We will select $T_\mathrm{max} = 270$ K and $T_\mathrm{min} = 175$ K in accordance with \citet{KS11}.\footnote{The lower limit on $T_P$ may be further extended if one takes additional greenhouse gases such as molecular hydrogen (H$_2$) into account \citep{PG11,RAF19}.} We solve for the times at which these temperatures are obtained and take their difference to obtain $t_\mathrm{HZ}$. After simplification, we end up with
\begin{equation}\label{tHab}
   t_\mathrm{HZ} \approx 3 \times 10^{-7}\,\mathrm{Gyr}\, \left(\frac{M_\mathrm{BD}}{M_J}\right)^{2.037} \left(\frac{a}{1\,\mathrm{AU}}\right)^{-1.543}.
\end{equation}
Upon inverting this equation and solving for the values of $a$ compatible with a given $t_\mathrm{HZ}$, we find
\begin{equation}
    a \approx 5.9 \times 10^{-5}\,\mathrm{AU}\,\left(\frac{M_\mathrm{BD}}{M_J}\right)^{1.32}\left(\frac{t_\mathrm{HZ}}{1\,\mathrm{Gyr}}\right)^{-0.648}.
\end{equation}

\begin{figure}
\includegraphics[width=7.5cm]{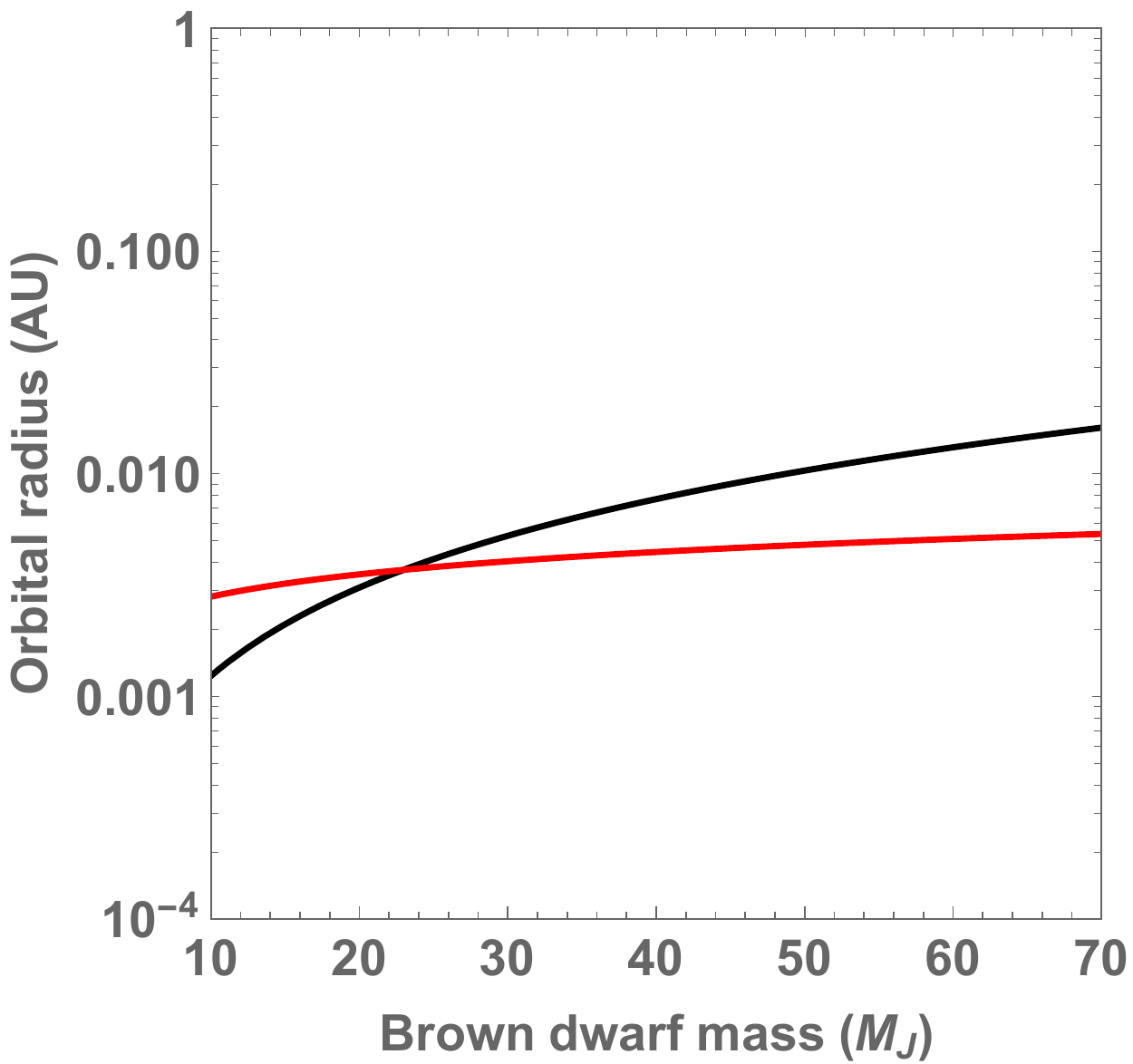} \\
\caption{The black curve shows the range of orbital radii (in AU) compatible with a $\sim 1$ Gyr habitability lifetime. It is plotted as a function of the brown dwarf mass (in $M_J$). The red curve depicts the Roche limit as a function of the brown dwarf mass.}
\label{FigBDHabOrb}
\end{figure}

We have plotted the compatible orbital radii for a given $t_\mathrm{HZ}$ in Fig. \ref{FigBDHZ}. The regions inside the curves and above the associated red lines (Roche limits) yield the compatible values of $a$. First, we observe that $t_\mathrm{HZ}$ increases as $a$ is lowered and vice-versa. Thus, the maximum lifetime occurs when the planet is situated close to its Roche limit. Second, we see that $t_\mathrm{HZ}$ increases with the mass of the brown dwarf; this is along expected lines because more massive brown dwarfs have higher luminosities. These results are, for the most part, in agreement with prior analyses of this topic \citep{AS04,BRL11,Barnes13}. For example, we find that the maximum value of $t_\mathrm{HZ}$ for a $70 M_J$ brown dwarf is $\sim 6$ Gyr, whereas \citet{AS04} obtained $4$-$10$ Gyr. Similarly, \citet{BRL11} found that brown dwarfs with masses $> 50 M_J$ could have maximum habitable zone lifetimes of $1$-$10$ Gyr, which happens to be consistent with our findings. 

By combining (\ref{RocFin}) and (\ref{tHab}), as well as inspecting Fig. \ref{FigBDHZ}, another interesting result is obtained. For $t_\mathrm{HZ} > 10$ Gyr to be valid at the Roche limit, we would require $M_\mathrm{BD} \gtrsim 100\,M_J$ (equivalent to $\gtrsim 0.1\,M_\odot$). This limit is higher than the conventional upper bound of $\sim 90\,M_J$ for brown dwarfs \citep{ZHP17}, although it has recently been demonstrated that brown dwarfs with masses up to $\sim 0.12\,M_\odot$ are theoretically feasible \citep{Forbes19}. Hence, our analysis suggests that brown dwarfs are less likely to host long-lived habitable zones than M-dwarfs because planets can remain in the habitable zones of late-type M-dwarfs for $\gtrsim 100$ Gyr \citep{Rushby13}. Thus, in principle, it would appear as though planets around brown dwarfs do not constitute targets as promising as M-dwarf exoplanets. This expectation should, however, be balanced by the fact that planets around M-dwarfs might be rendered inhospitable to life via intense stellar winds, space weather events and flares \citep{ManLo}.

As noted earlier, most major evolutionary breakthroughs on Earth required $\gtrsim 1$ Gyr after its formation \citep{JMS95,DW16,KN17}. In habitability studies, it is generally advantageous to retain habitable conditions over timescales of gigayears. Hence, by setting $t_\mathrm{HZ} \sim 1$ Gyr, we can investigate what range of orbital radii allow for this possibility as a function of $M_\mathrm{BD}$. The results are presented in Fig. \ref{FigBDHabOrb}.  The parameter space of interest is situated above the red curve and below the black curve. When the brown dwarf has a mass smaller than $\sim 20 M_J$, we find that it cannot remain in the habitable zone over an interval of $\sim 1$ Gyr. Hence, it is unlikely that such low-mass brown dwarfs host planets that remain habitable over geologically significant timescales.  

\section{Photosynthesis on Earth-like planets around brown dwarfs}\label{SecPHZ}
As a brown dwarf cools over time, it will emit radiation at longer wavelengths. Hence, the flux of photons received at shorter wavelengths drops dramatically (due to the exponential cutoff). Thus, the length of time over which an Earth-like planet receives sufficient photon flux for photosynthesis is constrained by this cooling. We will work out the salient details herein.

In our subsequent analysis, we shall adopt the conservative choice of oxygenic photosynthesis that uses two coupled photosystems and operates at wavelengths of $\lambda_\mathrm{min} = 400$ nm and $\lambda_\mathrm{max} = 750$ nm. While longer wavelengths are feasible in principle \citep{WoRa02,KST07,Lingam19}, no concrete empirical evidence is available thus far to substantiate this hypothesis. At wavelengths $\lesssim 400$ nm, it is likely that these photosystems would be subject to damage by ultraviolet (UV) radiation \citep{CoAi}. We will also adopt an ``Earth-like'' atmosphere that permits the majority of photosynthetically active radiation (PAR) to reach the surface, i.e., the corresponding optical depth is smaller than unity.

With these simplifications, the PAR flux incident on the planet ($\Phi_\mathrm{PAR}$) is given by
\begin{equation}\label{PhFlux}
    \Phi_\mathrm{PAR} \approx \frac{\dot{N}_\mathrm{BD}}{4\pi a^2}.
\end{equation}
Here, $\dot{N}_\mathrm{BD}$ is the photon flux emitted by the brown dwarf, which is defined as
\begin{equation}\label{NBDBlack}
\dot{N}_\star = 4 \pi R_\mathrm{BD}^2 \int_{\lambda_\mathrm{min}}^{\lambda_\mathrm{max}} \frac{2c}{\lambda^4}\left[\exp\left(\frac{h c}{\lambda k_B T_\mathrm{BD}}\right)-1\right]^{-1}\,d\lambda,
\end{equation}
for a black body spectrum.\footnote{The black body assumption is indubitably an idealization. However, as far as flares and other intermittent bursts are concerned, it is unlikely that these phenomena would affect our results significantly unless their frequency of occurrence is extremely high \citep{LiLo19}.} In this equation, note that $R_\mathrm{BD}$ and $T_\mathrm{BD}$ are given by (\ref{RadBD}) and (\ref{TeffBD}), respectively. Thus, we shall substitute (\ref{TeffBD}), (\ref{RadBD}) and (\ref{NBDBlack}) into (\ref{PhFlux}). After simplifying the resultant expression, we find that $\Phi_\mathrm{PAR}$ is expressible as
\begin{eqnarray}\label{FluxFin}
&& \Phi_\mathrm{PAR} \approx 9.5 \times 10^{13}\,\mathrm{m^{-2}\,s^{-1}} \left(\frac{a}{1\,\mathrm{AU}}\right)^{-2}  \\
&& \hspace{0.3in} \times \left(\frac{M_\mathrm{BD}}{M_J}\right)^{1.814}\left(\frac{t_\mathrm{BD}}{1\,\mathrm{Gyr}}\right)^{-0.972} \mathcal{F}\left(t_\mathrm{BD},M_\mathrm{BD}\right), \nonumber
\end{eqnarray}
where the function $\mathcal{F}\left(t_\mathrm{BD},M_\mathrm{BD}\right)$ is defined to be
\begin{equation}
\mathcal{F}\left(t_\mathrm{BD},M_\mathrm{BD}\right) \approx \int_{Y_1}^{Y_2} \frac{y'^2\,dy'}{\exp\left(y'\right) - 1}, 
\end{equation}
and the two limits of integration ($Y_1$ and $Y_2$) depend on $t_\mathrm{BD}$ and $M_\mathrm{BD}$ as described below:
\begin{eqnarray}
&& Y_1 \approx 610.3 \left(\frac{t_\mathrm{BD}}{1\,\mathrm{Gyr}}\right)^{0.324} \left(\frac{M_\mathrm{BD}}{M_J}\right)^{-0.827}, \nonumber \\
&& Y_2 \approx 325.5 \left(\frac{t_\mathrm{BD}}{1\,\mathrm{Gyr}}\right)^{0.324} \left(\frac{M_\mathrm{BD}}{M_J}\right)^{-0.827}.
\end{eqnarray}

A theoretical analysis by \citet{RKB00} (see also \citealt{WoRa02}) concluded that the minimum PAR flux necessary for photosynthesis based on biophysical constraints is $\Phi_c \approx 1.2 \times 10^{16}$ m$^{-2}$ s$^{-1}$. Thus, as the brown dwarf cools, the flux received by the planet will decline until it finally drops below $\Phi_c$ after a certain duration; here, we implicitly assume that the PAR flux immediately after planet formation is greater than $\Phi_c$. If the interval over which $\Phi_\mathrm{PAR} > \Phi_c$ holds true is taken to be the ``photosynthesis-zone lifetime'' ($t_\mathrm{PZ}$), we can derive a relationship between this quantity, $M_\mathrm{BD}$ and $a$ by setting $\Phi_\mathrm{PAR} = \Phi_c$. Thus, by using (\ref{FluxFin}) in conjunction with this data, we arrive at
\begin{equation}\label{aPhoComp}
    a = 0.09\,\mathrm{AU}\,\sqrt{\mathcal{F}(t_\mathrm{PZ},M_\mathrm{BD})}\left(\frac{M_\mathrm{BD}}{M_J}\right)^{0.907}\left(\frac{t_\mathrm{PZ}}{1\,\mathrm{Gyr}}\right)^{-0.486}.
\end{equation}
Thus, for given values of $a$ and $M_\mathrm{BD}$, we can solve for the corresponding $t_\mathrm{PZ}$. Alternatively, by choosing a range of values for $t_\mathrm{PZ}$ and $M_\mathrm{BD}$, we can investigate what choices of $a$ are compatible with these values. We will pursue the latter approach, whereby the permitted orbital radii ($a$) are estimated for different choices of $t_\mathrm{PZ}$ and $M_\mathrm{BD}$.

\begin{figure}
\includegraphics[width=7.5cm]{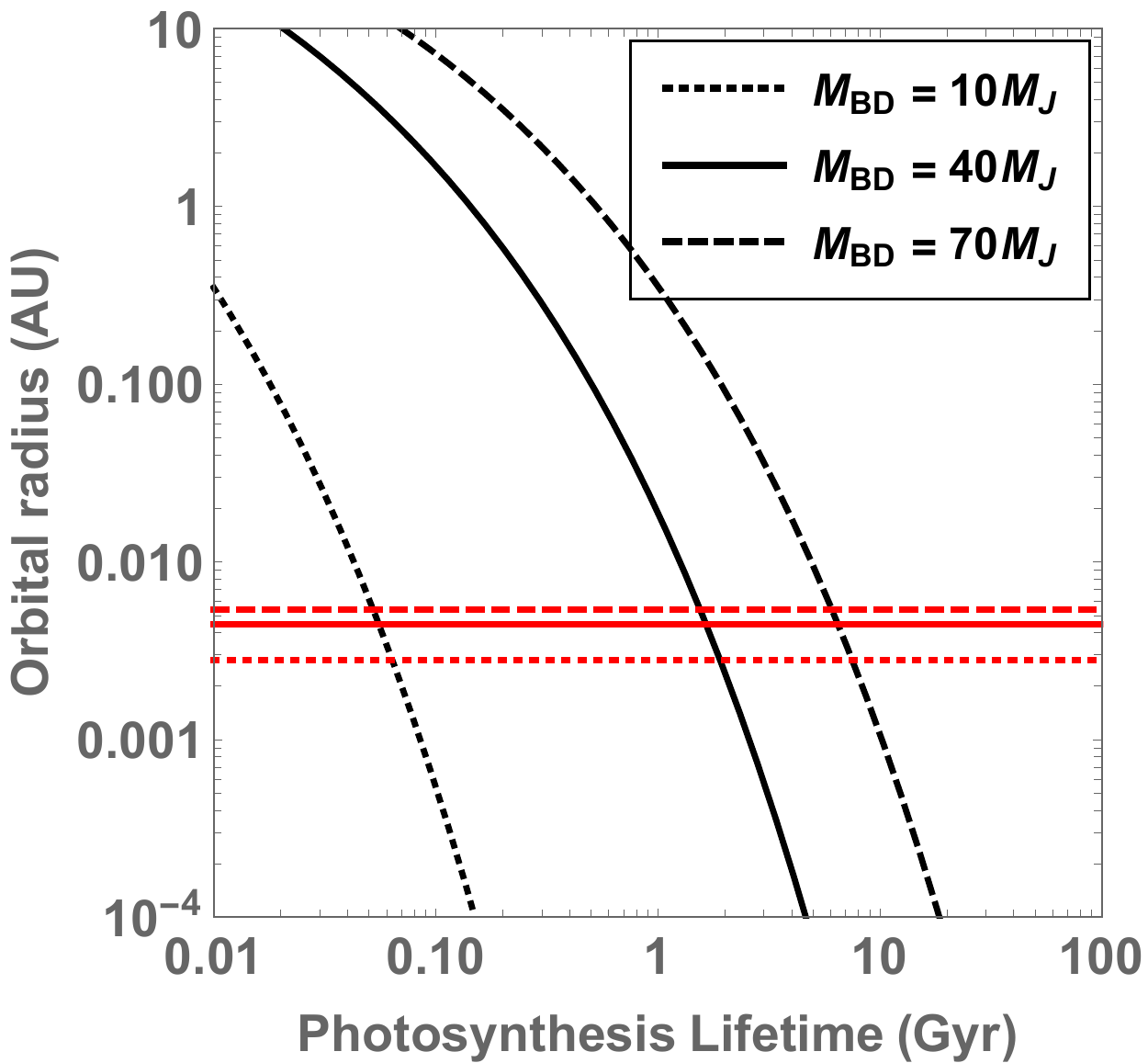} \\
\caption{The orbital radius (in AU) compatible with a given choice of the ``photosynthesis lifetime'' (in Gyr) - the interval over which an Earth-like planet receives sufficient PAR flux for photosynthesis - is plotted for brown dwarfs of different masses. The dotted, unbroken and dashed black curves correspond to brown dwarf masses of $10 M_J$, $40 M_J$ and $70 M_J$, respectively. The dotted, unbroken and dashed red lines are the Roche limits for $10 M_J$, $40 M_J$ and $70 M_J$ brown dwarfs, respectively.}
\label{FigPLife}
\end{figure}

We restrict ourselves to those brown dwarf-planet systems wherein the length of time that the planet spends in the photosynthetic ``zone'' is $> 10$ Myr for the same reasons outlined when tackling $t_\mathrm{HZ}$ in Sec. \ref{SecHZBD}. We also impose the additional constraint that $a$ must be larger than the Roche limit to avoid tidal disruption, as explained in Sec. \ref{SecHZBD}. We have plotted the compatible orbital radii as a function of $t_\mathrm{PZ}$ for different choices of the brown dwarf mass in Fig. \ref{FigPLife}. All values of the orbital radii lying inside the curves and upwards of the red lines (Roche limits) are permitted for a given choice of $t_\mathrm{PZ}$.

\begin{figure}
\includegraphics[width=7.5cm]{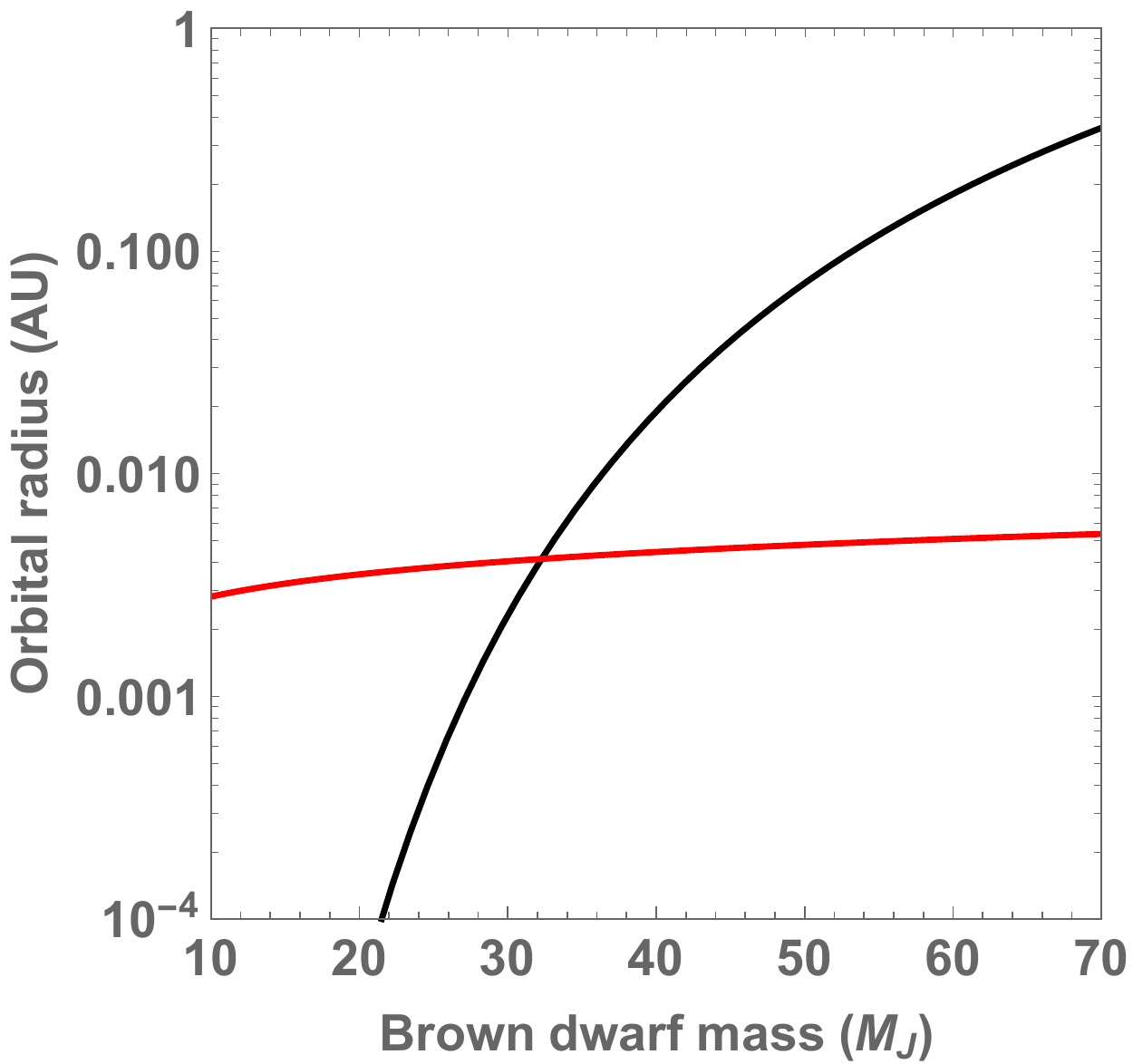} \\
\caption{The black curve shows the range of orbital radii (in AU) compatible with a $\sim 1$ Gyr photosynthesis lifetime is plotted as a function of the brown dwarf mass (in $M_J$). The red curve depicts the Roche limit as a function of the brown dwarf mass. The parameter space of interest lies above the red curve and below the black curve.}
\label{FigBDMOrb}
\end{figure}

There are some general conclusions that can be drawn from Fig. \ref{FigPLife}. First, as $a$ is decreased, we find that $t_\mathrm{PZ}$ increases. This is along expected lines because the planet will receive higher photon fluxes for a longer period of time. Second, for a fixed $a$, we see that $t_\mathrm{PZ}$ falls off rapidly with $M_\mathrm{BD}$. This is also consistent with intuition because smaller brown dwarfs possess a lower luminosity and will therefore emit sufficient PAR for a shorter duration. Lastly, we note that planets around a $10 M_J$ brown dwarf can never reside within the photosynthesis zone for $\gtrsim 1$ Gyr for reasons elucidated below.

If we set $t_\mathrm{PZ} \sim 1$ Gyr (see Sec. \ref{SecHZBD}), we can study what ranges of $a$ and $M_\mathrm{BD}$ enable this criterion to be satisfied. The result is plotted in Fig. \ref{FigBDMOrb}. The region below the black curve but above the red curve corresponds to the parameter space that permits the photosynthesis lifetime to satisfy $t_\mathrm{PZ} \gtrsim 1$ Gyr. One of the most striking results we find is that brown dwarfs with $M_\mathrm{BD} < 30 M_J$ do not fulfill the requisite criterion. Hence, it is unlikely that planets around these brown dwarfs are capable of sustaining photosynthetic biospheres over timescales of gigayears.

\begin{figure}
\includegraphics[width=7.5cm]{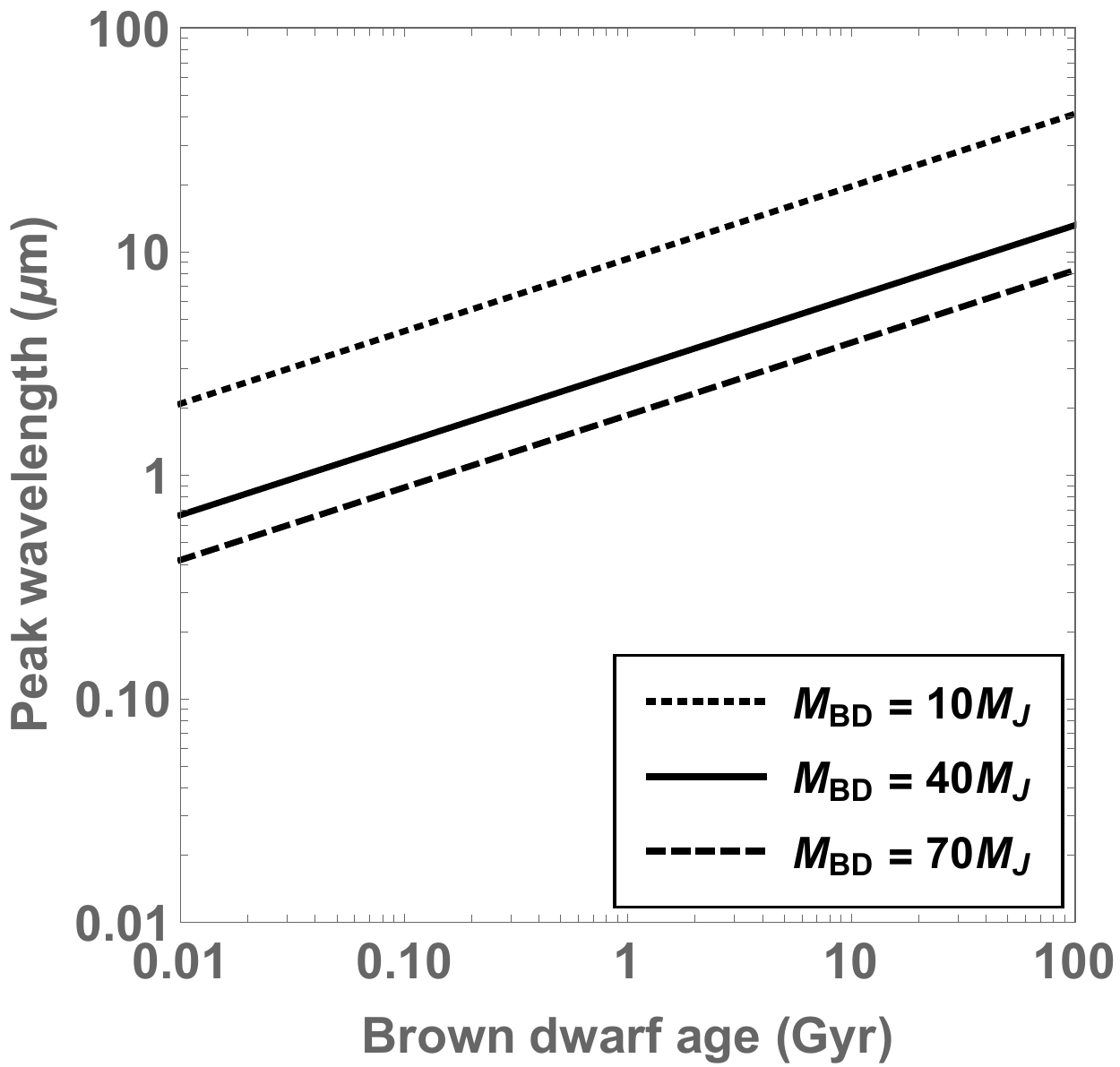} \\
\caption{The wavelength $\lambda_\mathrm{peak}$ (in $\mu$m) at which the maximal value of the brown dwarf's photon flux density occurs is plotted as a function of its age. The dotted, unbroken and dashed black curves depict $\lambda_\mathrm{peak}$ for brown dwarfs whose masses are $10 M_J$, $40 M_J$ and $70 M_J$, respectively.}
\label{FigBPeakWav}
\end{figure}

Before moving ahead, there are some important caveats that need to be reiterated. First, even in the context of oxygenic photosynthesis, it is theoretically feasible to extend the upper wavelength of $\sim 750$ nm to the near- and mid-infrared by coupling together a number of photosystems in series. This would presumably lower the quantum yield and may cause (or require) a proliferation of side reactions, but such coupled multi-photon schemes are feasible \citep{WoRa02}. It was conjectured by \citet{KS07a} that the absorbance peak of photopigments might be manifested at the peak incident photon flux. The latter, in turn, depends on both the spectral properties of the brown dwarf (or star) and the atmospheric composition of the planet. As planets can vary widely in composition, it is instructive to focus on the former, specifically the peak of the brown dwarf's photon flux density. By modelling this quantity as a black body, we find that the peak wavelength ($\lambda_\mathrm{peak}$) is given by
\begin{equation}\label{Abspeak}
 \lambda_\mathrm{peak} \approx 62.3\,\mathrm{\mu m}\,   \left(\frac{t_\mathrm{BD}}{1\,\mathrm{Gyr}}\right)^{0.324} \left(\frac{M_\mathrm{BD}}{M_J}\right)^{-0.827}
\end{equation}
In the case of the Sun, we find that $\lambda_\mathrm{peak} \approx 635$ nm, whereas the absorbance peak of the chlorophylls in photosystems I and II is $\sim 680$-$700$ nm \citep{KS07a}. It is thus evident that this simple ansatz might enable us to roughly gauge where the peak absorbance of photopigments could occur, and the approximate wavelength at which the spectral edge of photosynthetic organisms is manifested. In Fig. \ref{FigBPeakWav}, we have plotted $\lambda_\mathrm{peak}$ as a function of the brown dwarf's age for different masses. We observe that the peak shifts toward longer wavelengths for older and/or low-mass brown dwarfs. In most cases, we find that $\lambda_\mathrm{peak}$ has a range of $\sim 1$-$10$ $\mu$m, thus implying that the spectral edge may shift to near- and mid-infrared wavelengths over time. 

If we move beyond oxygenic photosynthesis, there are hypothesized alternatives such as ``hydrogenic'' photosynthesis which might function at wavelengths as long as $\sim 1.5$ $\mu$m \citep{BSZ14}. And last, but not least, even if photoautotrophs cannot survive, a diverse range of chemotrophs could exist on these planets. There are, in fact, a number of evolutionary features shared by chemoautotrophs and photoautotrophs, which has motivated several papers to discuss the possibility that the latter evolved from the former in appropriate geological environments \citep{MBB17,OES19} (see also \citealt{MaLi19}). On Earth, chemotrophs are predicted to make up $\sim 10\%$ of the total biomass \citep{BPM18}. Hence, it is therefore possible that planets around brown dwarfs might host thriving microbial biospheres even in the absence of sufficient PAR for photosynthesis.

\section{Ultraviolet radiation: prebiotic chemistry and abiogenesis}\label{SecUVAbio}

\begin{figure}
\includegraphics[width=7.5cm]{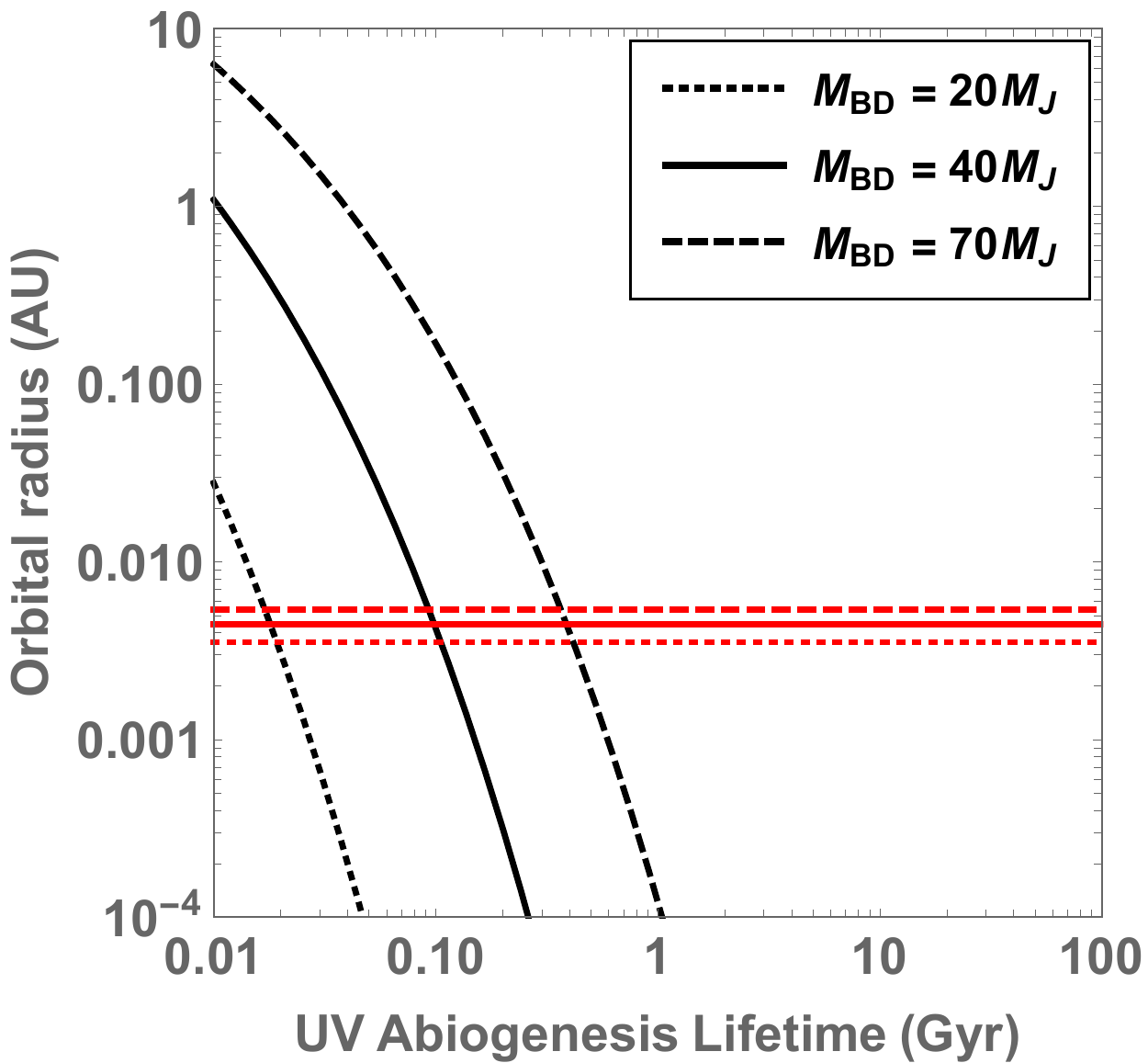} \\
\caption{The orbital radius (in AU) compatible with a given UV abiogenesis lifetime (in Gyr) - the interval over which an Earth-like planet receives sufficient UV flux for prebiotic chemistry leading to abiogenesis - is plotted for brown dwarfs of different masses. The dotted, unbroken and dashed black curves correspond to brown dwarf masses of $20 M_J$, $40 M_J$ and $70 M_J$, respectively. The dotted, unbroken and dashed red lines are the Roche limits for $20 M_J$, $40 M_J$ and $70 M_J$ brown dwarfs, respectively.}
\label{FigUVAbio}
\end{figure}

The concepts of the habitable zone and photosynthesis are fairly well understood - although some unknowns do exist - and are often regarded as potentially generic features in the search for extraterrestrial life. In contrast, the origin of life remains an arguably bigger mystery, and a number of competing hypotheses exist based on geochemical, physiological, and genomic considerations, to name a few \citep{Lu16,SM16,Walk17}. In this part of the paper we shall focus on a \emph{particular} proposal, with the express understanding that a number of other origin-of-life scenarios exist. In fact, as per recent models, even the possibility of some planets being seeded with life via galactic-scale panspermia ought not to be dismissed outright \citep{Ling16,LL18,GLL18}.

The proposal that we deal with is sometimes referred to as ``cyanosulfidic metabolism'' due to its reliance on hydrogen cyanide as a core feedstock molecule for synthesizing the precursors of major biomolecular building blocks such as lipids and amino acids \citep{PPR15,Suth16,Suth17}. These pathways have been the subject of numerous sophisticated laboratory experiments and require UV radiation in the range $200 < \lambda < 280$ nm. In order for the prebiotic reactions driven by UV photochemistry to function effectively, they must be more dominant than other ``dark'' reactions that unfold in the absence of UV radiation. In turn, this necessitates a minimum flux of UV photons in this range, whose value is $5.44 \times 10^{16}$ photons m$^{-2}$ s$^{-1}$ \citep{RXT18}; see also \citet{LGB19}.

With this new critical flux, it is fairly straightforward to analyze the brown dwarf-planet system along the lines of Sec. \ref{SecPHZ}. By doing so, we can determine the relationship between $a$, $M_\mathrm{BD}$ and the length of time in the ``abiogenesis zone" ($t_\mathrm{AZ}$); the latter quantifies the length of time during which the planet receives enough photons to permit the synthesis of biomolecular building blocks (and thence biopolymers) via UV-mediated reactions. As the steps are quite similar, we provide the final expression for the compatible values of $a$. The analog of (\ref{aPhoComp}) is therefore given by
\begin{equation}\label{aAbioComp}
    a = 0.04\,\mathrm{AU}\,\sqrt{\mathcal{G}(t_\mathrm{AZ},M_\mathrm{BD})}\left(\frac{M_\mathrm{BD}}{M_J}\right)^{0.907}\left(\frac{t_\mathrm{AZ}}{1\,\mathrm{Gyr}}\right)^{-0.486},
\end{equation}
where $\mathcal{G}(t_\mathrm{AZ},M_\mathrm{BD})$ is defined as
\begin{equation}
\mathcal{G}\left(t_\mathrm{AZ},M_\mathrm{BD}\right) \approx \int_{Z_1}^{Z_2} \frac{z'^2\,dz'}{\exp\left(z'\right) - 1}, 
\end{equation}
where the upper and lower limits of integration are
\begin{eqnarray}
&& Z_1 \approx 1220.5 \left(\frac{t_\mathrm{AZ}}{1\,\mathrm{Gyr}}\right)^{0.324} \left(\frac{M_\mathrm{BD}}{M_J}\right)^{-0.827}, \nonumber \\
&& Z_2 \approx 871.8 \left(\frac{t_\mathrm{AZ}}{1\,\mathrm{Gyr}}\right)^{0.324} \left(\frac{M_\mathrm{BD}}{M_J}\right)^{-0.827}.
\end{eqnarray}
In deriving these relations, we have implicitly assumed that the young planet in question is optically thin to UV radiation within the specified range; at the minimum, this requires the planet to resemble Hadean-Archean Earth in not having a sizable ozone layer.

\begin{figure}
\includegraphics[width=7.5cm]{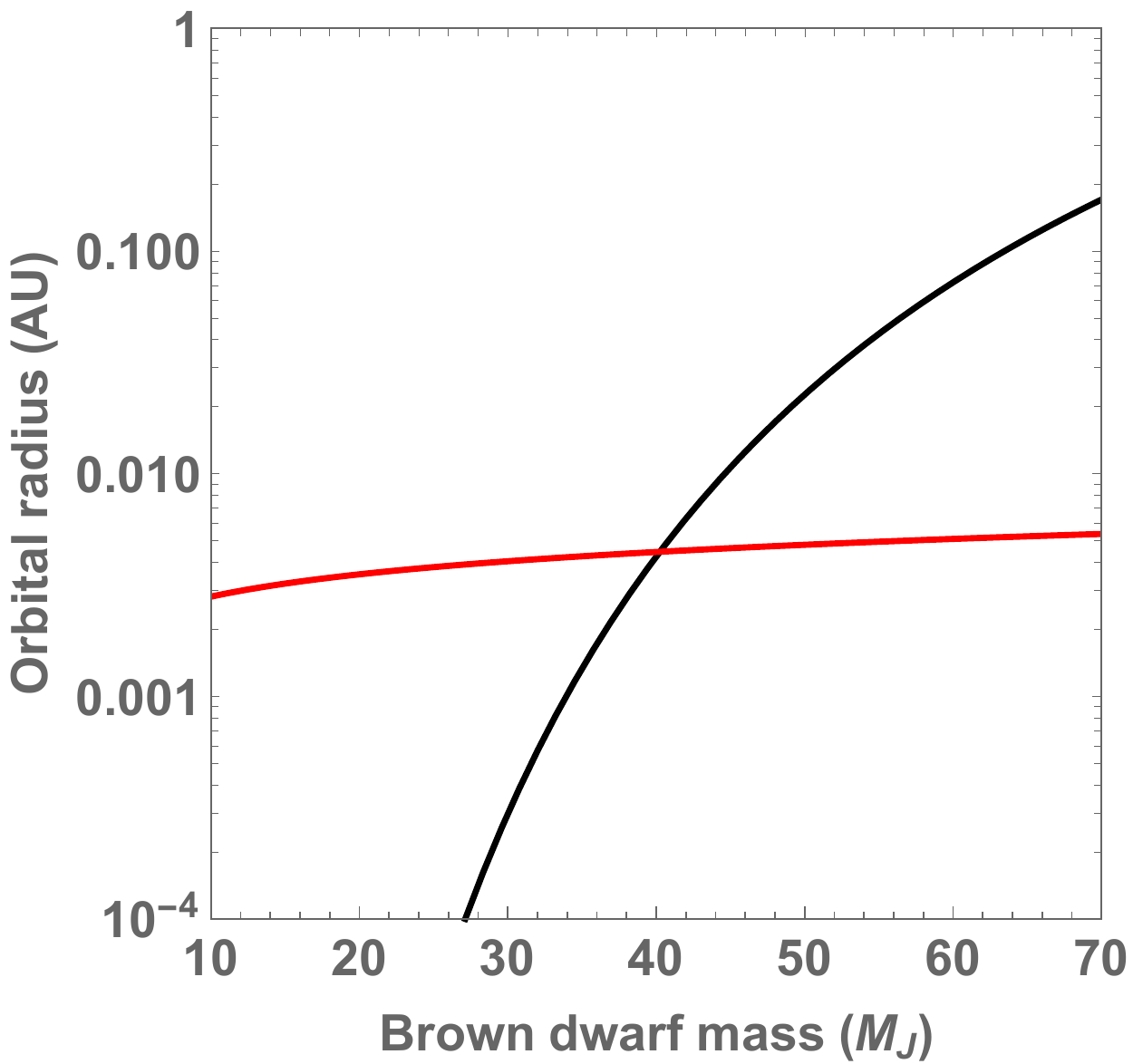} \\
\caption{The black curve depicts the range of orbital radii (in AU) compatible with a $\sim 100$ Myr lifetime for UV-mediated prebiotic chemistry and abiogenesis is plotted as a function of the brown dwarf mass (in $M_J$). The red curve depicts the Roche limit as a function of the brown dwarf mass. The parameter space in question spans the region above the red curve and below the black curve.}
\label{BDUVLim}
\end{figure}

From this equation, we can determine the values of $a$ compatible with a given $t_\mathrm{AZ}$. We also impose the dual constraints of $t_\mathrm{AZ} > 10$ Myr and $a > d_\mathrm{RL}$ (where $d_\mathrm{RL}$ is the Roche limit) for reasons explicated previously. The final result is depicted in Fig. \ref{FigUVAbio}. Note that we have plotted the curve for a $20 M_J$ brown dwarf instead of its $10 M_J$ counterpart because the latter does not fulfill the above two inequalities. From the plot, we see that $t_\mathrm{HZ}$ is higher when the orbital radius is smaller and when the brown dwarf mass is higher, both of which are consistent with expectations. For all brown dwarfs within the mass range considered in the paper, we discover that the two conditions $t_\mathrm{AZ} > 1$ Gyr and $a > d_\mathrm{RL}$ are not simultaneously realizable. 

We do not have a precise estimate for the time taken for life to originate on Earth. There is also the added complication that the abiogenesis timescale is not necessarily the same on other worlds \citep{ST12}. For the sake of further quantitative analysis, we will suppose that the transition from pre-life to life on other worlds occurs on a timescale comparable to that required on Earth. While the earliest definitive evidence for life on Earth dates from $\sim 3.7$ Ga, a combination of genomic and fossil analyses as well as theoretical arguments suggest that the timescale life's emergence on Earth ranged between $\mathcal{O}\left(10^7\right)$ and $\mathcal{O}\left(10^8\right)$ yrs at the most \citep{LM94,DPG17,BPC18}.

Thus, by erring on the side of caution, we shall specify $t_\mathrm{AZ} \sim 100$ Myr. By substituting this number into (\ref{aAbioComp}), we can determine what values of $a$ are compatible with this timescale as a function of $M_\mathrm{BD}$. In Fig. \ref{BDUVLim}, the zone of interest is depicted and it lies amid the black and red curves. By inspecting the plot, we find that the two curves intersect at $\sim 40 M_J$. Hence, as per our analysis, it is relatively unlikely for planets around brown dwarfs with masses $< 40 M_J$ to support UV-mediated prebiotic reactions over a significant timespan ($\gtrsim 100$ Myr). 

At this stage, it is worth reiterating that a dearth of UV radiation ought not be regarded as a death knell insofar as abiogenesis is concerned. As noted earlier, there are numerous hypotheses for the origin of life, of which one of the most well known posits that life originated at alkaline hydrothermal vents on the seafloor \citep{BH85,MB08,SHW16}. Clearly, in this scenario, the access to substantial fluxes of UV photons is unlikely to be a limiting factor. On account of this reason, we recommend that the results herein should be viewed with due caution. 

\section{Discussion and Conclusions}\label{SecDisc}
It is unknown exactly how many brown dwarfs reside within the Milky Way. However, recent surveys indicate that the corresponding number is potentially comparable to the number of stars \citep{Muz19}. Thus, in the most optimal circumstances, brown dwarfs might sustain as much life (on terrestrial planets) as stars.

To this end, we have studied how the habitability of Earth-like planets is affected by the brown dwarfs they orbit. We focused our attention on three different concepts of central importance in astrobiology: the habitable zone, photosynthesis and prebiotic chemistry leading to abiogenesis. To keep our analysis as conservative as possible, we restricted our attention to biochemical pathways and mechanisms found on Earth and to planets where the habitability intervals associated with each of these processes are comparable to those present on Earth. We took a number of parameters, such as the planet's orbital radius and the brown dwarf's mass and age, into account to develop the appropriate models.

Our basic conclusion is that planets around brown dwarfs with masses smaller than $\sim 20 M_J$, $\sim 30 M_J$ and $\sim 40 M_J$ have a low likelihood of sustaining long-term temperate climates (i.e., remaining in the habitable zone), photosynthesis and UV-mediated abiogenesis, respectively. Even when these limits are exceeded, the prospects for long-term habitability are determined by the orbital radius at which the planet is situated. Planets that are too far out ($\gtrsim 0.1$ AU) are not guaranteed to receive sufficient fluxes of electromagnetic radiation in the appropriate range, whereas overly close-in planets ($\lesssim 10^{-3}$ AU) are susceptible to tidal heating, desiccation and tidal disruption. Another crucial point that emerged from our analysis is that brown dwarfs have temporally limited habitable zones ($\lesssim 10$ Gyr) with respect to M-dwarfs; all other factors held equal, one would therefore not expect brown dwarfs to host habitable planets when their ages exceed $\sim 10$ Gyr.\footnote{Under the simplifying assumption of a uniform formation rate of brown dwarfs over cosmic time, one might expect $\sim 10\%$ of all brown dwarfs with $M_\mathrm{BD} \gtrsim 30 M_J$ to host ``active'' habitable zones.}

Recent statistical studies indicate that the number of brown dwarfs with $M_\mathrm{BD} > 30 M_J$ is potentially as high as $\sim 10^{11}$ because the ratio of main-sequence stars to brown dwarfs is apparently $\sim 2$-$5$ \citep{SGC13,Muzic17,Muz19,SDR19}; note that the number of stars in the Milky Way is $\gtrsim 10^{11}$. However, it is not possible to estimate the number of Earth-sized planets in the habitable zones of brown dwarfs ($\eta_\oplus$) at this stage due to the paucity of available statistics; in fact, even for stars, the uncertainty in the range of values for $\eta_\oplus$ approaches an order of magnitude \citep{Kal17}. However, in the event that $\eta_\oplus$ for brown dwarfs is comparable to that of stars, we see that the number of potentially habitable planets around brown dwarfs in the Milky Way might be similar to the number of such planets orbiting main-sequence stars. 

This raises the question of what types of biosignatures can be produced. To answer this question, recall that oxygenic photosynthesis may be feasible over timescales of gigayears provided that $M_\mathrm{BD} > 30 M_J$. Thus, planets around these brown dwarfs might evolve oxygenic photosynthesis. The real issue, however, is that the net primary productivity on these worlds will be photon-limited in addition to other constraints imposed by access to nutrients and water.\footnote{A recent quantitative analysis by \citet{LL2019} (see also \citealt{WP13}) suggests that worlds with too much or too little liquid water on the surface are potentially unlikely to accumulate atmospheric oxygen to appreciable levels.} Drawing an analogy with M-dwarf exoplanets \citep{LCP18,LiLo19}, it is possible that planets around brown dwarfs could likewise possess anoxic atmospheres (with negligible ozone layers) despite the presence of biospheres, and could thereby give rise to ``false negatives'' during searches for biological O$_2$ and O$_3$.

\begin{figure}
\includegraphics[width=7.5cm]{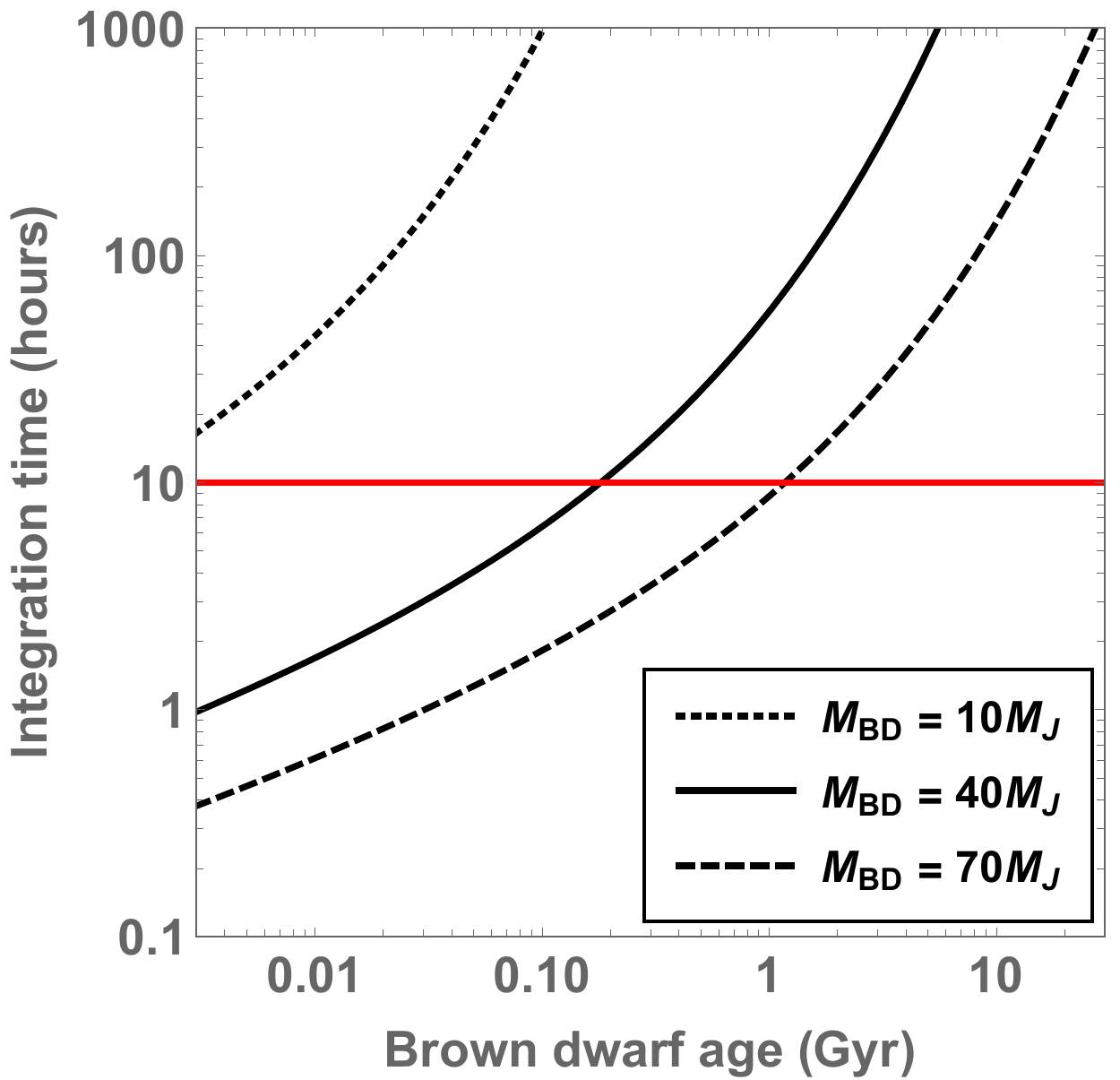} \\
\caption{The JWST integration time ($\Delta t$) necessary to achieve a SNR of $\sim 5$ using transmission spectroscopy for an Earth-analog around a brown dwarf is depicted as a function of the brown dwarf's age. We have set $d = 10$ pc and $\lambda = 3$ $\mu$m in (\ref{TSSNR}). The dotted, unbroken and dashed black curves depict $\Delta t$ for brown dwarfs whose masses are $10 M_J$, $40 M_J$ and $70 M_J$, respectively. The horizontal red line demarcates the desirable region where the integration time is $< 10$ hrs.}
\label{FigTSIntT}
\end{figure}

Oxygenic photosynthesis not only gives rise to O$_2$ but also to the ``vegetation red edge'' at $\sim 0.7$ $\mu$m \citep{STSF}, which may be discernible in principle by means of photometric observations provided that a sufficient fraction of the surface is covered by vegetation. On the other hand, if photosynthesis in the infrared is feasible, it is theoretically conceivable that the peak absorbance of photopigments might evolve over time in accordance with (\ref{Abspeak}). Thus, once a brown dwarf's mass and age are known, we can use (\ref{Abspeak}) to determine the approximate wavelength at which the spectral edge might exist. There are a number of other atmospheric and surface biosignatures arising from autotrophs and heterotrophs; a comprehensive review of this subject can be found in \citet{SKP18}.

The next question that arises has to do with the relative ease of detecting biosignatures. We will draw upon the scaling relations presented in \citet{FAD18} for our purpose. For starters, we note that the transit depth and transit probability for Earth-like planets around brown dwarfs are of the order of $1\%$, both of which are higher than (or similar to) the estimates for planets around main-sequence stars. If we opt to carry out transmission spectroscopy for an Earth-analog using the JWST,\footnote{\url{https://www.jwst.nasa.gov/}} the resultant signal-to-noise ratio (SNR), after invoking \citet{FAD18}, is expressible as:
\begin{eqnarray}\label{TSSNR}
 &&   \mathrm{SNR} \sim 3\left(\frac{M_\mathrm{BD}}{M_J}\right)^{1/3} \left(\frac{d}{10\,\mathrm{pc}}\right)^{-1}\left(\frac{\Delta t}{30\,\mathrm{hr}}\right)^{1/2} \nonumber \\
 && \hspace{0.5in} \times \left(\frac{\dot{n}\left(\lambda; T_\mathrm{BD} \right)}{\dot{n}(3\,\mathrm{\mu m}; 2500\,\mathrm{K})}\right)^{1/2},
\end{eqnarray}
where $\Delta t$ is the integration time, $d$ is the distance to the brown dwarf from Earth, and $\dot{n}\left(\lambda; T_\mathrm{BD} \right)$ is the photon flux density for a black body at temperature $T_\mathrm{BD}$ and measured at wavelength $\lambda$; it equals the integrand present in (\ref{NBDBlack}).  If we desire a SNR of $\sim 5$, we can invert the above equation to solve for $\Delta t$ as a function of the brown dwarf's age and mass. The resultant integration time has been plotted in Fig. \ref{FigTSIntT}. By inspecting this figure, we see that an integration time of $\mathcal{O}(10)$ hr is achievable only for comparatively young ($\sim 0.1$-$1$ Gyr) and massive ($M_\mathrm{BD} \gtrsim 40 M_J$) brown dwarfs. 

\begin{figure}
\includegraphics[width=7.5cm]{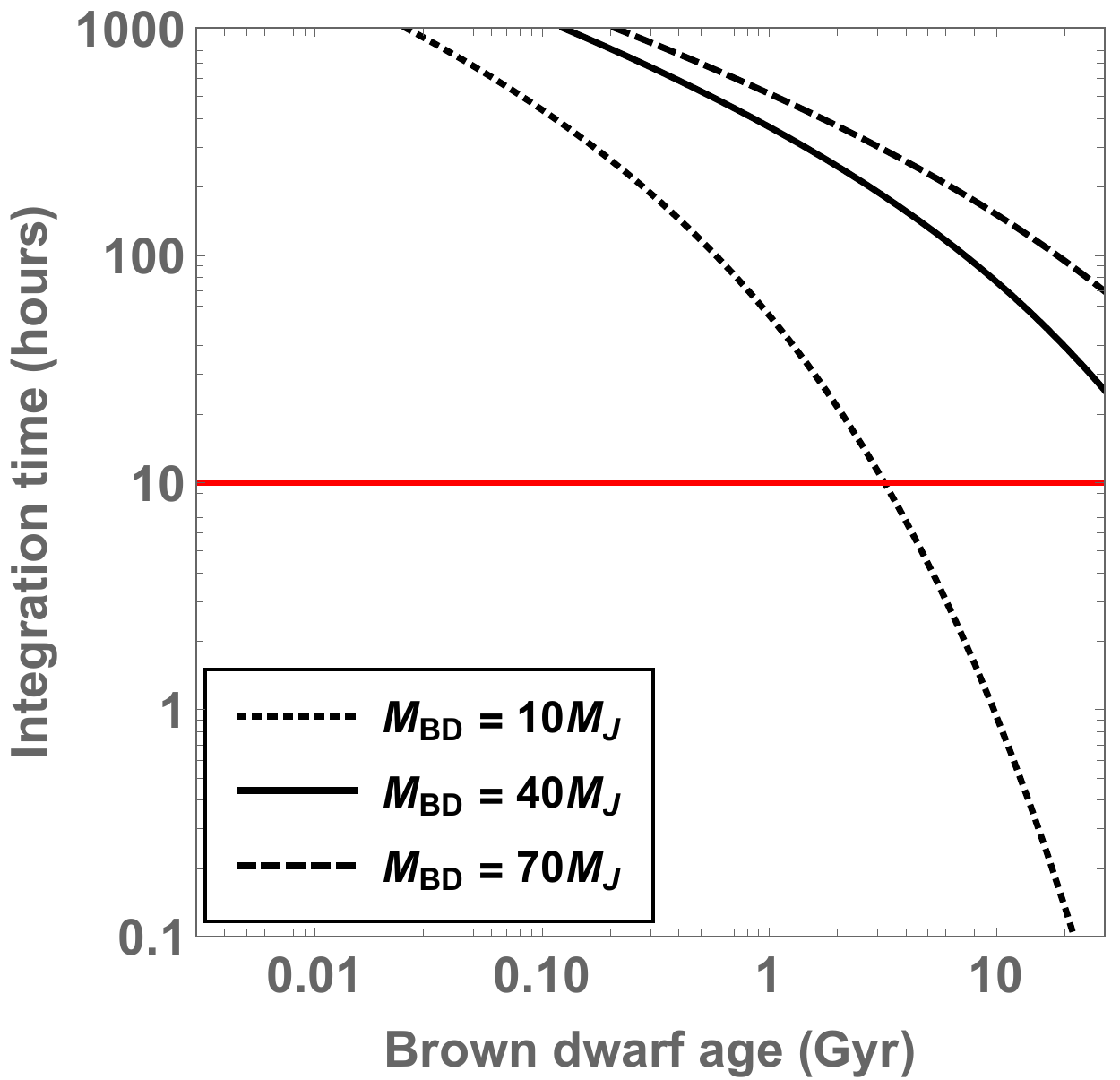} \\
\caption{The JWST integration time ($\Delta t$) necessary to achieve a SNR of $\sim 5$ using eclipse (i.e., thermal emission) spectroscopy for an Earth-analog around a brown dwarf is depicted as a function of the brown dwarf's age. We have specified $d = 10$ pc, $\lambda = 10$ $\mu$m and $\zeta = 0.5$ in (\ref{TESNR}). The dotted, unbroken and dashed black curves depict $\Delta t$ for brown dwarfs whose masses are $10 M_J$, $40 M_J$ and $70 M_J$, respectively. The horizontal red line demarcates the desirable region where the integration time is $< 10$ hrs.}
\label{FigTEIntT}
\end{figure}

Alternatively, one can analyze the thermal emission of the planet by computing the spectra of the system during secondary eclipses and taking the difference. For this procedure, the SNR for an Earth-analog using the parameters of the JWST is \citep{FAD18}:
\begin{eqnarray}\label{TESNR}
 &&   \mathrm{SNR} \sim 0.5 \zeta \left(\frac{M_\mathrm{BD}}{M_J}\right)^{1/3} \left(\frac{d}{10\,\mathrm{pc}}\right)^{-1}\left(\frac{\Delta t}{30\,\mathrm{hr}}\right)^{1/2} \nonumber \\
 && \hspace{0.5in} \times \left(\frac{\dot{n}\left(\lambda; T_\mathrm{BD} \right)}{\dot{n}(10\,\mathrm{\mu m}; 2500\,\mathrm{K})}\right)^{-1/2},
\end{eqnarray}
where $\zeta$ represents the normalized depth of the spectral features generated by putative biosignatures. Note that the photon flux density of the brown dwarf is raised to a negative power, implying that older (and therefore cooler) and smaller brown dwarfs ought to be favored by this method. This intuition is borne out by Fig. \ref{FigTEIntT}, from which we find that thermal emission spectroscopy yields an integration time of $\mathcal{O}(10)$ hr for low-mass ($M_\mathrm{BD} \sim 10 M_J$) brown dwarfs that have ages of $\gtrsim 1$ Gyr.

Lastly, when it comes to direct imaging via scattered light, it is feasible to achieve contrast ratios of the order of $\sim 10^{-6}$ for Earth-analogs because the contrast between the planet and the substellar object in the visible and near-infrared ($\mathcal{C}$) is expressible as
\begin{equation}
    \mathcal{C} \sim 10^{-10} \left(\frac{a}{1\,\mathrm{AU}}\right)^{-2}.
\end{equation}
The primary obstacle, however, is that the small orbital radius of the planet, which increases the contrast as seen above, also leads to an angular separation between the planet and star that is lower than the inner working angle of current coronagraphs and starshades as well as charge injection devices \citep{BFB16}.

Finally, we wish to highlight some other crucial issues that merit further study. First, it is necessary to determine the typical composition of Earth-sized planets around brown dwarfs. There is tentative evidence suggesting that these planets might be water-poor and carbon-rich \citep{PHC13}; if correct, the former factor is detrimental to habitability insofar as life-as-we-know-it is concerned. Observations also indicate that the abundance of certain feedstock molecules - such as HCN, whose importance was described in Sec. \ref{SecUVAbio} - could be much lower in protoplanetary disks around brown dwarfs \citep{PAL09}. If prebiotic chemistry is dependent on the exogenous delivery of these compounds, it is plausible that exoplanets orbiting brown dwarfs would be hampered in this regard. 

Second, a sizable fraction of brown dwarfs are binaries, either in conjunction with other brown dwarfs or with stellar companions (see \citealt{RMC16,FBB18} and references therein). Although planets around binaries appear to possess some intrinsic advantages \citep{Shev17}, they are also potentially susceptible to other issues such as formation of terrestrial planets, orbital destabilization over geological timescales and rapid ($\sim 10^3$ yr) fluctuations in climate \citep{Cu14,JAP,Forg16,PLSA}; however, some of the issues alleged to suppress the habitability of certain circumbinary exoplanets may be overstated \citep{PE17}. Third, accurate dynamical treatments are necessary to determine whether lithopanspermia in multi-planet systems around brown dwarfs is several orders of magnitude more likely from a dynamical perspective when compared to the Earth-to-Mars transfer of rocky ejecta, as argued in \citet{LL17}.

Last and not least, the issues of atmospheric escape and ocean desiccation merit thorough scrutiny. While some studies have addressed these topics from the standpoint of X-rays and UV radiation \citep{Barnes13,Bol18}, it is being increasingly appreciated that stellar winds and coronal mass ejections (CMEs) may contribute significantly in both respects \citep{DL17,DL18,DHL,aira19}. It is therefore necessary to synthesize observational limits on brown dwarf magnetic fields \citep{KPW}, theoretical models and observational constraints for stellar winds and CMEs, and multi-fluid magnetohydrodynamic simulations to determine the rates of atmospheric escape and water depletion \citep{DH17,DLM18}. If theoretical models of M-dwarf exoplanets are anything to go by \citep{Lin18,ManLo}, it is conceivable that exoplanets around brown dwarfs might be likewise depleted of $\lesssim 10$ bar atmospheres in timescales less than $1$ Gyr. However, it is necessary to utilize models of volatile delivery by means of impactors in tandem \citep{MCM15,TGJ18,BL18}, because they can serve to replenish the water reservoirs of brown dwarf exoplanets. 

In summary, we investigated the prospects for long-term habitability of exoplanets around brown dwarfs by studying their habitable zones and prospects for photosynthesis and UV-mediated abiogenesis. We found that planets around brown dwarfs with masses $\lesssim 30 M_J$ are comparatively unlikely to be habitable (or inhabited) over timescales of gigayears. We also briefly explored the potential number of such brown dwarfs in our Galaxy, and the prospects for detecting biosignatures on temperate planets orbiting them. Our analysis suggests that there may be nearly as many astrobiological targets around brown dwarfs as around main-sequence stars under the most optimal circumstances, thereby highlighting the need for in-depth theoretical, experimental and observational studies in the future. 

\section*{Acknowledgements}

We thank Adam Burrows for helpful clarifications concerning the paper. We also thank the reviewer for the insightful report that helped improve the quality of the paper. This work was supported in part by the Breakthrough Prize Foundation, Harvard University, and the Institute for Theory and Computation. 


\begin{thebibliography}{}
\expandafter\ifx\csname natexlab\endcsname\relax\def\natexlab#1{#1}\fi
\providecommand{\url}[1]{\href{#1}{#1}}

\bibitem[{{Airapetian} {et~al.}(2019){Airapetian}, {Barnes}, {Cohen},
  {Collinson}, {Danchi}, {Dong}, {Del Genio}, {France}, {Garcia-Sage},
  {Glocer}, {Gopalswamy}, {Grenfell}, {Gronoff}, {G``udel}, {Herbst},
  {Henning}, {Jackman}, {Jin}, {Johnstone}, {Kaltenegger}, {Kay}, {Kobayashi},
  {Kuang}, {Li}, {Lynch}, {L``uftinger}, {Luhmann}, {Maehara}, {Mlynczak},
  {Notsu}, {Ramirez}, {Rugheimer}, {Scheucher}, {Schlieder}, {Shibata},
  {Sousa-Silva}, {Stamenkovi'c}, {Strangeway}, {Usmanov}, {Vergados},
  {Verkhoglyadova}, {Vidotto}, {Voytek}, {Way}, {Zank}, \&
  {Yamashiki}}]{aira19}
{Airapetian}, V.~S., {Barnes}, R., {Cohen}, O., {et~al.} 2019, Int. J.
  Astrobiol., doi:10.1017/S1473550419000132

\bibitem[{{Andreeshchev} \& {Scalo}(2004)}]{AS04}
{Andreeshchev}, A., \& {Scalo}, J. 2004, in IAU Symposium, Vol. 213,
  Bioastronomy 2002: Life Among the Stars, ed. R.~{Norris} \& F.~{Stootman},
  115--118

\bibitem[{{Apai}(2013)}]{Apai13}
{Apai}, D. 2013, Astronomische Nachrichten, 334, 57

\bibitem[{{Apai} {et~al.}(2005){Apai}, {Pascucci}, {Bouwman}, {Natta},
  {Henning}, \& {Dullemond}}]{Apai05}
{Apai}, D., {Pascucci}, I., {Bouwman}, J., {et~al.} 2005, Science, 310, 834

\bibitem[{{Auddy} {et~al.}(2016){Auddy}, {Basu}, \& {Valluri}}]{Auddy16}
{Auddy}, S., {Basu}, S., \& {Valluri}, S.~R. 2016, Advances in Astronomy, 2016,
  574327

\bibitem[{{Bains} {et~al.}(2014){Bains}, {Seager}, \& {Zsom}}]{BSZ14}
{Bains}, W., {Seager}, S., \& {Zsom}, A. 2014, Life, 4, 716

\bibitem[{{Bar-On} {et~al.}(2018){Bar-On}, {Phillips}, \& {Milo}}]{BPM18}
{Bar-On}, Y.~M., {Phillips}, R., \& {Milo}, R. 2018, Proc. Natl. Acad. Sci.
  USA, 115, 6506

\bibitem[{{Barnes} \& {Heller}(2013)}]{Barnes13}
{Barnes}, R., \& {Heller}, R. 2013, Astrobiology, 13, 279

\bibitem[{{Baross} \& {Hoffman}(1985)}]{BH85}
{Baross}, J.~A., \& {Hoffman}, S.~E. 1985, Orig. Life Evol. Biosph., 15, 327

\bibitem[{{Batcheldor} {et~al.}(2016){Batcheldor}, {Foadi}, {Bahr}, {Jenne},
  {Ninkov}, {Bhaskaran}, \& {Chapman}}]{BFB16}
{Batcheldor}, D., {Foadi}, R., {Bahr}, C., {et~al.} 2016, \pasp, 128, 025001

\bibitem[{{Betts} {et~al.}(2018){Betts}, {Puttick}, {Clark}, {Williams},
  {Donoghue}, \& {Pisani}}]{BPC18}
{Betts}, H.~C., {Puttick}, M.~N., {Clark}, J.~W., {et~al.} 2018, Nat. Ecol.
  Evol., 2, 1556

\bibitem[{{Bolmont}(2018)}]{Bol18}
{Bolmont}, E. 2018, {Habitability in Brown Dwarf Systems}, ed. H.~J. {Deeg} \&
  J.~A. {Belmonte} (Springer), 3069--3090

\bibitem[{{Bolmont} {et~al.}(2011){Bolmont}, {Raymond}, \& {Leconte}}]{BRL11}
{Bolmont}, E., {Raymond}, S.~N., \& {Leconte}, J. 2011, \aap, 535, A94

\bibitem[{{Brunini} \& {L{\'o}pez}(2018)}]{BL18}
{Brunini}, A., \& {L{\'o}pez}, M.~C. 2018, \mnras, 479, 1392

\bibitem[{{Burgasser} {et~al.}(2006){Burgasser}, {Burrows}, \&
  {Kirkpatrick}}]{Burgasser06}
{Burgasser}, A.~J., {Burrows}, A., \& {Kirkpatrick}, J.~D. 2006, \apj, 639,
  1095

\bibitem[{{Burgasser} {et~al.}(2002){Burgasser}, {Kirkpatrick}, {Brown},
  {Reid}, {Burrows}, {Liebert}, {Matthews}, {Gizis}, {Dahn}, {Monet}, {Cutri},
  \& {Skrutskie}}]{Burgasser02}
{Burgasser}, A.~J., {Kirkpatrick}, J.~D., {Brown}, M.~E., {et~al.} 2002, \apj,
  564, 421

\bibitem[{{Burrows} {et~al.}(2001){Burrows}, {Hubbard}, {Lunine}, \&
  {Liebert}}]{Burrows01}
{Burrows}, A., {Hubbard}, W.~B., {Lunine}, J.~I., \& {Liebert}, J. 2001,
  Reviews of Modern Physics, 73, 719

\bibitem[{{Burrows} \& {Liebert}(1993)}]{Burrows93}
{Burrows}, A., \& {Liebert}, J. 1993, Reviews of Modern Physics, 65, 301

\bibitem[{{Caballero}(2018)}]{Cab18}
{Caballero}, J.~A. 2018, Geosciences, 8, 362

\bibitem[{{Carter}(2008)}]{Cart08}
{Carter}, B. 2008, IJAsB, 7, 177

\bibitem[{{Chauvin} {et~al.}(2004){Chauvin}, {Lagrange}, {Dumas}, {Zuckerman},
  {Mouillet}, {Song}, {Beuzit}, \& {Lowrance}}]{Chauvin04}
{Chauvin}, G., {Lagrange}, A.~M., {Dumas}, C., {et~al.} 2004, \aap, 425, L29

\bibitem[{{Cockell} \& {Airo}(2002)}]{CoAi}
{Cockell}, C.~S., \& {Airo}, A. 2002, Origins of Life and Evolution of the
  Biosphere, 32, 255

\bibitem[{{Cuntz}(2014)}]{Cu14}
{Cuntz}, M. 2014, \apj, 780, 14

\bibitem[{{Cushing}(2014)}]{Cushing14}
{Cushing}, M.~C. 2014, in Astrophysics and Space Science Library, Vol. 401, 50
  Years of Brown Dwarfs, ed. V.~{Joergens}, 113

\bibitem[{{Cushing} {et~al.}(2011){Cushing}, {Kirkpatrick}, {Gelino},
  {Griffith}, {Skrutskie}, {Mainzer}, {Marsh}, {Beichman}, {Burgasser},
  {Prato}, {Simcoe}, {Marley}, {Saumon}, {Freedman}, {Eisenhardt}, \&
  {Wright}}]{Cushing11}
{Cushing}, M.~C., {Kirkpatrick}, J.~D., {Gelino}, C.~R., {et~al.} 2011, \apj,
  743, 50

\bibitem[{{Daemgen} {et~al.}(2016){Daemgen}, {Natta}, {Scholz}, {Testi},
  {Jayawardhana}, {Greaves}, \& {Eastwood}}]{DNS16}
{Daemgen}, S., {Natta}, A., {Scholz}, A., {et~al.} 2016, \aap, 594, A83

\bibitem[{{Dawkins} \& {Wong}(2016)}]{DW16}
{Dawkins}, R., \& {Wong}, Y. 2016, {The Ancestor's Tale: A Pilgrimage to the
  Dawn of Evolution}, 2nd edn. (Mariner Books)

\bibitem[{{Desidera}(1999)}]{Desi99}
{Desidera}, S. 1999, \pasp, 111, 1529

\bibitem[{{Dodd} {et~al.}(2017){Dodd}, {Papineau}, {Grenne}, {Slack},
  {Rittner}, {Pirajno}, {O'Neil}, \& {Little}}]{DPG17}
{Dodd}, M.~S., {Papineau}, D., {Grenne}, T., {et~al.} 2017, Nature, 543, 60

\bibitem[{{Dole}(1964)}]{Dole}
{Dole}, S.~H. 1964, {Habitable planets for man} (Blaisdell Pub.~Co.)

\bibitem[{{Dong} {et~al.}(2019){Dong}, {Huang}, \& {Lingam}}]{DHL}
{Dong}, C., {Huang}, Z., \& {Lingam}, M. 2019, \apjl, 882, L16

\bibitem[{{Dong} {et~al.}(2017{\natexlab{a}}){Dong}, {Huang}, {Lingam},
  {T{\'o}th}, {Gombosi}, \& {Bhattacharjee}}]{DH17}
{Dong}, C., {Huang}, Z., {Lingam}, M., {et~al.} 2017{\natexlab{a}}, Astrophys.
  J. Lett., 847, L4

\bibitem[{{Dong} {et~al.}(2018{\natexlab{a}}){Dong}, {Jin}, {Lingam},
  {Airapetian}, {Ma}, \& {van der Holst}}]{DL18}
{Dong}, C., {Jin}, M., {Lingam}, M., {et~al.} 2018{\natexlab{a}}, PNAS, 115,
  260

\bibitem[{{Dong} {et~al.}(2017{\natexlab{b}}){Dong}, {Lingam}, {Ma}, \&
  {Cohen}}]{DL17}
{Dong}, C., {Lingam}, M., {Ma}, Y., \& {Cohen}, O. 2017{\natexlab{b}}, \apjl,
  837, L26

\bibitem[{{Dong} {et~al.}(2018{\natexlab{b}}){Dong}, {Lee}, {Ma}, {Lingam},
  {Bougher}, {Luhmann}, {Curry}, {Toth}, {Nagy}, {Tenishev}, {Fang},
  {Mitchell}, {Brain}, \& {Jakosky}}]{DLM18}
{Dong}, C., {Lee}, Y., {Ma}, Y., {et~al.} 2018{\natexlab{b}}, \apjl, 859, L14

\bibitem[{{Fontanive} {et~al.}(2018){Fontanive}, {Biller}, {Bonavita}, \&
  {Allers}}]{FBB18}
{Fontanive}, C., {Biller}, B., {Bonavita}, M., \& {Allers}, K. 2018, Mon. Not.
  R. Astron. Soc., 479, 2702

\bibitem[{{Forbes} \& {Loeb}(2019)}]{Forbes19}
{Forbes}, J.~C., \& {Loeb}, A. 2019, \apj, 871, 227

\bibitem[{{Forgan}(2016)}]{Forg16}
{Forgan}, D. 2016, \mnras, 463, 2768

\bibitem[{{Fujii} {et~al.}(2018){Fujii}, {Angerhausen}, {Deitrick},
  {Domagal-Goldman}, {Grenfell}, {Hori}, {Kane}, {Pall{\'e}}, {Rauer},
  {Siegler}, {Stapelfeldt}, \& {Stevenson}}]{FAD18}
{Fujii}, Y., {Angerhausen}, D., {Deitrick}, R., {et~al.} 2018, Astrobiology,
  18, 739

\bibitem[{{Geballe} {et~al.}(2002){Geballe}, {Knapp}, {Leggett}, {Fan},
  {Golimowski}, {Anderson}, {Brinkmann}, {Csabai}, {Gunn}, {Hawley},
  {Hennessy}, {Henry}, {Hill}, {Hindsley}, {Ivezi{\'c}}, {Lupton}, {McDaniel},
  {Munn}, {Narayanan}, {Peng}, {Pier}, {Rockosi}, {Schneider}, {Smith},
  {Strauss}, {Tsvetanov}, {Uomoto}, {York}, \& {Zheng}}]{Geballe02}
{Geballe}, T.~R., {Knapp}, G.~R., {Leggett}, S.~K., {et~al.} 2002, \apj, 564,
  466

\bibitem[{{Ginsburg} {et~al.}(2018){Ginsburg}, {Lingam}, \& {Loeb}}]{GLL18}
{Ginsburg}, I., {Lingam}, M., \& {Loeb}, A. 2018, \apjl, 868, L12

\bibitem[{{Guo} {et~al.}(2010){Guo}, {Zhang}, {Zhang}, \& {Han}}]{GZZ10}
{Guo}, J., {Zhang}, F., {Zhang}, X., \& {Han}, Z. 2010, Astrophys. Space Sci.,
  325, 25

\bibitem[{{Han} {et~al.}(2016){Han}, {Bennett}, {Udalski}, \& {Jung}}]{HBU16}
{Han}, C., {Bennett}, D.~P., {Udalski}, A., \& {Jung}, Y.~K. 2016, \apj, 825, 8

\bibitem[{{Han} {et~al.}(2013){Han}, {Jung}, {Udalski}, {Sumi}, {Gaudi},
  {Gould}, {Bennett}, {Tsapras}, {Szyma{\'n}ski}, {Kubiak}, {Pietrzy{\'n}ski},
  {Soszy{\'n}ski}, {Skowron}, {Koz{\l}owski}, {Poleski}, {Ulaczyk},
  {Wyrzykowski}, {Pietrukowicz}, {OGLE Collaboration}, {Abe}, {Bond},
  {Botzler}, {Chote}, {Freeman}, {Fukui}, {Furusawa}, {Harris}, {Itow}, {Ling},
  {Masuda}, {Matsubara}, {Muraki}, {Ohnishi}, {Rattenbury}, {Saito},
  {Sullivan}, {Sweatman}, {Suzuki}, {Tristram}, {Wada}, {Yock}, {MOA
  Collaboration}, {Batista}, {Christie}, {Choi}, {DePoy}, {Dong}, {Hwang},
  {Kavka}, {Lee}, {Monard}, {Natusch}, {Ngan}, {Park}, {Pogge}, {Porritt},
  {Shin}, {Tan}, {Yee}, {{\ensuremath{\mu}}FUN Collaboration}, {Alsubai},
  {Bozza}, {Bramich}, {Browne}, {Dominik}, {Horne}, {Hundertmark}, {Ipatov},
  {Kains}, {Liebig}, {Snodgrass}, {Steele}, {Street}, \& {RoboNet
  Collaboration}}]{Han13}
{Han}, C., {Jung}, Y.~K., {Udalski}, A., {et~al.} 2013, \apj, 778, 38

\bibitem[{{Hayashi} \& {Nakano}(1963)}]{Hayashi63}
{Hayashi}, C., \& {Nakano}, T. 1963, Progress of Theoretical Physics, 30, 460

\bibitem[{{Jaime} {et~al.}(2014){Jaime}, {Aguilar}, \& {Pichardo}}]{JAP}
{Jaime}, L.~G., {Aguilar}, L., \& {Pichardo}, B. 2014, \mnras, 443, 260

\bibitem[{{Jung} {et~al.}(2018){Jung}, {Udalski}, {Gould}, {Ryu}, {Yee}, {and},
  {Han}, {Albrow}, {Lee}, {Kim}, {Hwang}, {Chung}, {Shin}, {Zhu}, {Cha}, {Kim},
  {Lee}, {Park}, {Lee}, {Kim}, {Pogge}, {KMTNet Collaboration},
  {Szyma{\'n}ski}, {Mr{\'o}z}, {Poleski}, {Skowron}, {Pietrukowicz},
  {Soszy{\'n}ski}, {Koz{\l}owski}, {Ulaczyk}, {Pawlak}, {Rybicki}, \& {OGLE
  Collaboration}}]{Jung18}
{Jung}, Y.~K., {Udalski}, A., {Gould}, A., {et~al.} 2018, \aj, 155, 219

\bibitem[{{Kaltenegger}(2017)}]{Kal17}
{Kaltenegger}, L. 2017, Annu. Rev. Astron. Astrophys., 55, 433

\bibitem[{{Kaltenegger} \& {Sasselov}(2011)}]{KS11}
{Kaltenegger}, L., \& {Sasselov}, D. 2011, \apjl, 736, L25

\bibitem[{{Kao} {et~al.}(2019){Kao}, {Pineda}, {Williams}, {Yadav}, {Shulyak},
  {Saur}, {Stevenson}, {Schmidt}, {Burgasser}, {Hallinan}, \& {Cruz}}]{KPW}
{Kao}, M., {Pineda}, J.~S., {Williams}, P., {et~al.} 2019, in Bulletin of the
  American Astronomical Society, Vol.~51, 484

\bibitem[{{Kasting} {et~al.}(1993){Kasting}, {Whitmire}, \&
  {Reynolds}}]{Kasting:93}
{Kasting}, J.~F., {Whitmire}, D.~P., \& {Reynolds}, R.~T. 1993, \icarus, 101,
  108

\bibitem[{{Kiang} {et~al.}(2007{\natexlab{a}}){Kiang}, {Siefert}, {Govindjee},
  \& {Blankenship}}]{KS07a}
{Kiang}, N.~Y., {Siefert}, J., {Govindjee}, \& {Blankenship}, R.~E.
  2007{\natexlab{a}}, Astrobiology, 7, 222

\bibitem[{{Kiang} {et~al.}(2007{\natexlab{b}}){Kiang}, {Segura}, {Tinetti},
  {Govindjee}, {Blankenship}, {Cohen}, {Siefert}, {Crisp}, \&
  {Meadows}}]{KST07}
{Kiang}, N.~Y., {Segura}, A., {Tinetti}, G., {et~al.} 2007{\natexlab{b}},
  Astrobiology, 7, 252

\bibitem[{{Kirkpatrick}(2005)}]{Kirk05}
{Kirkpatrick}, J.~D. 2005, \araa, 43, 195

\bibitem[{{Kirkpatrick} {et~al.}(1991){Kirkpatrick}, {Henry}, \&
  {McCarthy}}]{Kirkpatrick91}
{Kirkpatrick}, J.~D., {Henry}, T.~J., \& {McCarthy}, Donald~W., J. 1991, \apjs,
  77, 417

\bibitem[{{Kirkpatrick} {et~al.}(1999){Kirkpatrick}, {Reid}, {Liebert},
  {Cutri}, {Nelson}, {Beichman}, {Dahn}, {Monet}, {Gizis}, \&
  {Skrutskie}}]{Kirkpatrick99}
{Kirkpatrick}, J.~D., {Reid}, I.~N., {Liebert}, J., {et~al.} 1999, \apj, 519,
  802

\bibitem[{{Kirkpatrick} {et~al.}(2012){Kirkpatrick}, {Gelino}, {Cushing},
  {Mace}, {Griffith}, {Skrutskie}, {Marsh}, {Wright}, {Eisenhardt}, {McLean},
  {Mainzer}, {Burgasser}, {Tinney}, {Parker}, \& {Salter}}]{KGC12}
{Kirkpatrick}, J.~D., {Gelino}, C.~R., {Cushing}, M.~C., {et~al.} 2012, \apj,
  753, 156

\bibitem[{{Kirkpatrick} {et~al.}(2019){Kirkpatrick}, {Martin}, {Smart},
  {Cayago}, {Beichman}, {Marocco}, {Gelino}, {Faherty}, {Cushing}, {Schneider},
  {Mace}, {Tinney}, {Wright}, {Lowrance}, {Ingalls}, {Vrba}, {Munn}, {Dahm}, \&
  {McLean}}]{KMS19}
{Kirkpatrick}, J.~D., {Martin}, E.~C., {Smart}, R.~L., {et~al.} 2019, \apjs,
  240, 19

\bibitem[{{Knoll}(2015)}]{Knoll15}
{Knoll}, A.~H. 2015, {Life on a Young Planet: The First Three Billion Years of
  Evolution on Earth}, Princeton Science Library (Princeton University Press)

\bibitem[{{Knoll} \& {Nowak}(2017)}]{KN17}
{Knoll}, A.~H., \& {Nowak}, M.~A. 2017, Sci. Adv., 3, e1603076

\bibitem[{{Kopparapu} {et~al.}(2013){Kopparapu}, {Ramirez}, {Kasting}, {Eymet},
  {Robinson}, {Mahadevan}, {Terrien}, {Domagal-Goldman}, {Meadows}, \&
  {Deshpande}}]{Kopparapu13}
{Kopparapu}, R.~K., {Ramirez}, R., {Kasting}, J.~F., {et~al.} 2013, \apj, 765,
  131

\bibitem[{{Kumar}(1962{\natexlab{a}})}]{SSK62}
{Kumar}, S.~S. 1962{\natexlab{a}}, \aj, 67, 579

\bibitem[{{Kumar}(1962{\natexlab{b}})}]{Kum62}
---. 1962{\natexlab{b}}, {Models for Stars of Very Low Mass}, Tech. rep.

\bibitem[{{Kumar}(1963)}]{Kumar63}
---. 1963, \apj, 137, 1121

\bibitem[{{Lazcano} \& {Miller}(1994)}]{LM94}
{Lazcano}, A., \& {Miller}, S.~L. 1994, J. Mol. Evol., 39, 546

\bibitem[{{Lehmer} {et~al.}(2018){Lehmer}, {Catling}, {Parenteau}, \&
  {Hoehler}}]{LCP18}
{Lehmer}, O.~R., {Catling}, D.~C., {Parenteau}, M.~N., \& {Hoehler}, T.~M.
  2018, \apj, 859, 171

\bibitem[{{Lingam}(2016)}]{Ling16}
{Lingam}, M. 2016, \mnras, 455, 2792

\bibitem[{{Lingam} {et~al.}(2019){Lingam}, {Ginsburg}, \& {Bialy}}]{LGB19}
{Lingam}, M., {Ginsburg}, I., \& {Bialy}, S. 2019, \apj, 877, 62

\bibitem[{{Lingam} \& {Loeb}(2017)}]{LL17}
{Lingam}, M., \& {Loeb}, A. 2017, PNAS, 114, 6689

\bibitem[{{Lingam} \& {Loeb}(2018{\natexlab{a}})}]{LL18}
---. 2018{\natexlab{a}}, \aj, 156, 193

\bibitem[{{Lingam} \& {Loeb}(2018{\natexlab{b}})}]{Lin18}
---. 2018{\natexlab{b}}, Int. J. Astrobiol., 17, 116

\bibitem[{{Lingam} \& {Loeb}(2019{\natexlab{a}})}]{Lingam19}
---. 2019{\natexlab{a}}, Astrophys. J., 883, 143

\bibitem[{{Lingam} \& {Loeb}(2019{\natexlab{b}})}]{ManLo}
---. 2019{\natexlab{b}}, Rev. Mod. Phys., 91, 021002

\bibitem[{{Lingam} \& {Loeb}(2019{\natexlab{c}})}]{LL19}
---. 2019{\natexlab{c}}, Int. J. Astrobiol., 18, 527

\bibitem[{{Lingam} \& {Loeb}(2019{\natexlab{d}})}]{LiLo19}
---. 2019{\natexlab{d}}, \mnras, 485, 5924

\bibitem[{{Lingam} \& {Loeb}(2019{\natexlab{e}})}]{MaLi19}
---. 2019{\natexlab{e}}, Int. J. Astrobiol., doi:10.1017/S1473550419000247

\bibitem[{{Lingam} \& {Loeb}(2019{\natexlab{f}})}]{LL2019}
---. 2019{\natexlab{f}}, \aj, 157, 25

\bibitem[{{Lodders}(2010)}]{Lod10}
{Lodders}, K. 2010, {Exoplanet Chemistry}, ed. R.~{Barnes} (Wiley‐VCH),
  157--186

\bibitem[{{Luhman}(2012)}]{Luh12}
{Luhman}, K.~L. 2012, \araa, 50, 65

\bibitem[{{Luhman}(2014)}]{Luh14}
---. 2014, \apjl, 786, L18

\bibitem[{{Luisi}(2016)}]{Lu16}
{Luisi}, P.~L. 2016, The Emergence of Life: From Chemical Origins to Synthetic
  Biology (Cambridge Univ. Press)

\bibitem[{{Martin} {et~al.}(1997){Martin}, {Basri}, {Delfosse}, \&
  {Forveille}}]{Martin97}
{Martin}, E.~L., {Basri}, G., {Delfosse}, X., \& {Forveille}, T. 1997, \aap,
  327, L29

\bibitem[{{Martin} {et~al.}(2008){Martin}, {Baross}, {Kelley}, \&
  {Russell}}]{MB08}
{Martin}, W., {Baross}, J., {Kelley}, D., \& {Russell}, M.~J. 2008, Nat. Rev.
  Microbiol., 6, 805

\bibitem[{{Martin} {et~al.}(2017){Martin}, {Bryant}, \& {Beatty}}]{MBB17}
{Martin}, W.~F., {Bryant}, D.~A., \& {Beatty}, J.~T. 2017, FEMS Microbiol.
  Rev., 42, 205

\bibitem[{{Mulders} {et~al.}(2015){Mulders}, {Ciesla}, {Min}, \&
  {Pascucci}}]{MCM15}
{Mulders}, G.~D., {Ciesla}, F.~J., {Min}, M., \& {Pascucci}, I. 2015, \apj,
  807, 9

\bibitem[{{Murray} \& {Dermott}(1999)}]{MD99}
{Murray}, C.~D., \& {Dermott}, S.~F. 1999, {Solar system dynamics} (Cambridge
  University Press)

\bibitem[{{Mu{\v z}i{\'c}} {et~al.}(2019){Mu{\v z}i{\'c}}, {Scholz}, {Pe{\~n}a
  Ram{\'{\i}}rez}, {Jayawardhana}, {Sch{\"o}del}, {Geers}, {Cieza}, \&
  {Bayo}}]{Muz19}
{Mu{\v z}i{\'c}}, K., {Scholz}, A., {Pe{\~n}a Ram{\'{\i}}rez}, K., {et~al.}
  2019, \apj, 881, 79

\bibitem[{{Mu{\v{z}}i{\'c}} {et~al.}(2017){Mu{\v{z}}i{\'c}}, {Sch{\"o}del},
  {Scholz}, {Geers}, {Jayawardhana}, {Ascenso}, \& {Cieza}}]{Muzic17}
{Mu{\v{z}}i{\'c}}, K., {Sch{\"o}del}, R., {Scholz}, A., {et~al.} 2017, \mnras,
  471, 3699

\bibitem[{{Onstott} {et~al.}(2019){Onstott}, {Ehlmann}, {Sapers}, {Coleman},
  {Ivarsson}, {Marlow}, {Neubeck}, \& {Niles}}]{OES19}
{Onstott}, T.~C., {Ehlmann}, B.~L., {Sapers}, H., {et~al.} 2019, Astrobiology,
  19, 1230

\bibitem[{{Oppenheimer} {et~al.}(1995){Oppenheimer}, {Kulkarni}, {Matthews}, \&
  {Nakajima}}]{Oppenheimer95}
{Oppenheimer}, B.~R., {Kulkarni}, S.~R., {Matthews}, K., \& {Nakajima}, T.
  1995, Science, 270, 1478

\bibitem[{{Pascucci} {et~al.}(2009){Pascucci}, {Apai}, {Luhman}, {Henning},
  {Bouwman}, {Meyer}, {Lahuis}, \& {Natta}}]{PAL09}
{Pascucci}, I., {Apai}, D., {Luhman}, K., {et~al.} 2009, \apj, 696, 143

\bibitem[{{Pascucci} {et~al.}(2013){Pascucci}, {Herczeg}, {Carr}, \&
  {Bruderer}}]{PHC13}
{Pascucci}, I., {Herczeg}, G., {Carr}, J.~S., \& {Bruderer}, S. 2013, \apj,
  779, 178

\bibitem[{{Patel} {et~al.}(2015){Patel}, {Percivalle}, {Ritson}, {Duffy}, \&
  {Sutherland}}]{PPR15}
{Patel}, B.~H., {Percivalle}, C., {Ritson}, D.~J., {Duffy}, C.~D., \&
  {Sutherland}, J.~D. 2015, Nat. Chem., 7, 301

\bibitem[{{Payne} \& {Lodato}(2007)}]{Payne07}
{Payne}, M.~J., \& {Lodato}, G. 2007, \mnras, 381, 1597

\bibitem[{{Pierrehumbert} \& {Gaidos}(2011)}]{PG11}
{Pierrehumbert}, R., \& {Gaidos}, E. 2011, \apjl, 734, L13

\bibitem[{{Pilat-Lohinger} {et~al.}(2019){Pilat-Lohinger}, {Eggl}, \&
  {Bazs\'o}}]{PLSA}
{Pilat-Lohinger}, E., {Eggl}, S., \& {Bazs\'o}, A. 2019, Advances in Planetary
  Science, Vol.~4, {Planetary Habitability in Binary Systems} (World
  Scientific)

\bibitem[{{Popp} \& {Eggl}(2017)}]{PE17}
{Popp}, M., \& {Eggl}, S. 2017, Nat. Commun., 8, 14957

\bibitem[{{Ramirez} {et~al.}(2019){Ramirez}, {Abbot}, {Fujii}, {Hamano},
  {Kite}, {Levi}, {Lingam}, {Lueftinger}, {Robinson}, {Rushby}, {Schaefer},
  {Tasker}, {Vladilo}, \& {Wordsworth}}]{RAF19}
{Ramirez}, R., {Abbot}, D.~S., {Fujii}, Y., {et~al.} 2019, in Bulletin of the
  American Astronomical Society, Vol.~51, 31

\bibitem[{{Ramirez}(2018)}]{Ram18}
{Ramirez}, R.~M. 2018, Geosciences, 8, 280

\bibitem[{{Raven} \& {Donnelly}(2013)}]{RaD13}
{Raven}, J.~A., \& {Donnelly}, S. 2013, in {Habitability of Other Planets and
  Satellites}, ed. J.-P. {de Vera} \& J.~{Seckbach} (Springer), 267--284

\bibitem[{{Raven} {et~al.}(2000){Raven}, {K{\"u}bler}, \& {Beardall}}]{RKB00}
{Raven}, J.~A., {K{\"u}bler}, J.~E., \& {Beardall}, J. 2000, J. Mar. Biol.
  Assoc. UK, 80, 1

\bibitem[{{Rebolo}(2014)}]{Rebolo14}
{Rebolo}, R. 2014, in Astrophysics and Space Science Library, Vol. 401, 50
  Years of Brown Dwarfs, ed. V.~{Joergens}, 25

\bibitem[{{Rebolo} {et~al.}(1995){Rebolo}, {Zapatero Osorio}, \&
  {Mart{\'\i}n}}]{Rebolo95}
{Rebolo}, R., {Zapatero Osorio}, M.~R., \& {Mart{\'\i}n}, E.~L. 1995, \nat,
  377, 129

\bibitem[{{Reggiani} {et~al.}(2016){Reggiani}, {Meyer}, {Chauvin}, {Vigan},
  {Quanz}, {Biller}, {Bonavita}, {Desidera}, {Delorme}, {Hagelberg}, {Maire},
  {Boccaletti}, {Beuzit}, {Buenzli}, {Carson}, {Covino}, {Feldt}, {Girard},
  {Gratton}, {Henning}, {Kasper}, {Lagrange}, {Mesa}, {Messina}, {Montagnier},
  {Mordasini}, {Mouillet}, {Schlieder}, {Segransan}, {Thalmann}, \&
  {Zurlo}}]{RMC16}
{Reggiani}, M., {Meyer}, M.~R., {Chauvin}, G., {et~al.} 2016, Astron.
  Astrophys., 586, A147

\bibitem[{{Rimmer} {et~al.}(2018){Rimmer}, {Xu}, {Thompson}, {Gillen},
  {Sutherland}, \& {Queloz}}]{RXT18}
{Rimmer}, P.~B., {Xu}, J., {Thompson}, S.~J., {et~al.} 2018, Sci. Adv., 4,
  eaar3302

\bibitem[{{Rushby} {et~al.}(2013){Rushby}, {Claire}, {Osborn}, \&
  {Watson}}]{Rushby13}
{Rushby}, A.~J., {Claire}, M.~W., {Osborn}, H., \& {Watson}, A.~J. 2013, AsBio,
  13, 833

\bibitem[{{Scholz} {et~al.}(2013){Scholz}, {Geers}, {Clark}, {Jayawardhana}, \&
  {Muzic}}]{SGC13}
{Scholz}, A., {Geers}, V., {Clark}, P., {Jayawardhana}, R., \& {Muzic}, K.
  2013, \apj, 775, 138

\bibitem[{{Schwieterman} {et~al.}(2018){Schwieterman}, {Kiang}, {Parenteau},
  {Harman}, {DasSarma}, {Fisher}, {Arney}, {Hartnett}, {Reinhard}, {Olson},
  {Meadows}, {Cockell}, {Walker}, {Grenfell}, {Hegde}, {Rugheimer}, {Hu}, \&
  {Lyons}}]{SKP18}
{Schwieterman}, E.~W., {Kiang}, N.~Y., {Parenteau}, M.~N., {et~al.} 2018,
  Astrobiology, 18, 663

\bibitem[{{Seager} {et~al.}(2005){Seager}, {Turner}, {Schafer}, \&
  {Ford}}]{STSF}
{Seager}, S., {Turner}, E.~L., {Schafer}, J., \& {Ford}, E.~B. 2005,
  Astrobiology, 5, 372

\bibitem[{{Shapley}(1967)}]{Shap67}
{Shapley}, H. 1967, {Beyond the Observatory} (Charles Scribner's Sons)

\bibitem[{{Shevchenko}(2017)}]{Shev17}
{Shevchenko}, I.~I. 2017, \aj, 153, 273

\bibitem[{{Smith} \& {Morowitz}(2016)}]{SM16}
{Smith}, E., \& {Morowitz}, H.~J. 2016, {The Origin and Nature of Life on
  Earth: The Emergence of the Fourth Geosphere} (Cambridge University Press)

\bibitem[{{Smith} \& {Szathmary}(1995)}]{JMS95}
{Smith}, J.~M., \& {Szathmary}, E. 1995, {The Major Transitions in Evolution}
  (Oxford University Press)

\bibitem[{{Sojo} {et~al.}(2016){Sojo}, {Herschy}, {Whicher}, {Camprub{\'{\i}}},
  \& {Lane}}]{SHW16}
{Sojo}, V., {Herschy}, B., {Whicher}, A., {Camprub{\'{\i}}}, E., \& {Lane}, N.
  2016, Astrobiology, 16, 181

\bibitem[{{Spiegel} {et~al.}(2011){Spiegel}, {Burrows}, \&
  {Milsom}}]{Spiegel11}
{Spiegel}, D.~S., {Burrows}, A., \& {Milsom}, J.~A. 2011, \apj, 727, 57

\bibitem[{{Spiegel} \& {Turner}(2012)}]{ST12}
{Spiegel}, D.~S., \& {Turner}, E.~L. 2012, Proc. Natl. Acad. Sci. USA, 109, 395

\bibitem[{{Su{\'a}rez} {et~al.}(2019){Su{\'a}rez}, {Downes},
  {Rom{\'a}n-Z{\'u}{\~n}iga}, {Cervi{\~n}o}, {Brice{\~n}o}, {Petr-Gotzens}, \&
  {Vivas}}]{SDR19}
{Su{\'a}rez}, G., {Downes}, J.~J., {Rom{\'a}n-Z{\'u}{\~n}iga}, C., {et~al.}
  2019, \mnras, 486, 1718

\bibitem[{{Sutherland}(2016)}]{Suth16}
{Sutherland}, J.~D. 2016, Angew. Chem. Int. Ed., 55, 104

\bibitem[{{Sutherland}(2017)}]{Suth17}
---. 2017, Nat. Rev. Chem., 1, 0012

\bibitem[{{Tian} {et~al.}(2018){Tian}, {G{\"u}del}, {Johnstone}, {Lammer},
  {Luger}, \& {Odert}}]{TGJ18}
{Tian}, F., {G{\"u}del}, M., {Johnstone}, C.~P., {et~al.} 2018, \ssr, 214, 65

\bibitem[{{Todorov} {et~al.}(2010){Todorov}, {Luhman}, \& {McLeod}}]{TLM10}
{Todorov}, K., {Luhman}, K.~L., \& {McLeod}, K.~K. 2010, \apjl, 714, L84

\bibitem[{{Udalski} {et~al.}(2015){Udalski}, {Jung}, {Han}, {Gould},
  {Koz{\l}owski}, {Skowron}, {Poleski}, {Soszy{\'n}ski}, {Pietrukowicz},
  {Mr{\'o}z}, {Szyma{\'n}ski}, {Wyrzykowski}, {Ulaczyk}, {Pietrzy{\'n}ski},
  {Shvartzvald}, {Maoz}, {Kaspi}, {Gaudi}, {Hwang}, {Choi}, {Shin}, {Park}, \&
  {Bozza}}]{Udalski15}
{Udalski}, A., {Jung}, Y.~K., {Han}, C., {et~al.} 2015, \apj, 812, 47

\bibitem[{{Walker}(2017)}]{Walk17}
{Walker}, S.~I. 2017, Rep. Prog. Phys., 80, 092601

\bibitem[{{Wolstencroft} \& {Raven}(2002)}]{WoRa02}
{Wolstencroft}, R.~D., \& {Raven}, J.~A. 2002, Icarus, 157, 535

\bibitem[{{Wordsworth} \& {Pierrehumbert}(2013)}]{WP13}
{Wordsworth}, R.~D., \& {Pierrehumbert}, R.~T. 2013, \apj, 778, 154

\bibitem[{{Yates} {et~al.}(2017){Yates}, {Palmer}, {Biller}, \&
  {Cockell}}]{Yates17}
{Yates}, J.~S., {Palmer}, P.~I., {Biller}, B., \& {Cockell}, C.~S. 2017, \apj,
  836, 184

\bibitem[{{Zhang} {et~al.}(2017){Zhang}, {Homeier}, {Pinfield}, {Lodieu},
  {Jones}, {Allard}, \& {Pavlenko}}]{ZHP17}
{Zhang}, Z.~H., {Homeier}, D., {Pinfield}, D.~J., {et~al.} 2017, Mon. Not. R.
  Astron. Soc., 468, 261

\end{thebibliography}

\end{document}